\documentclass[11pt]{article}
\usepackage{graphicx}
\usepackage{amssymb}
\usepackage{lscape}

\def\cer{Cherenkov }
\def\naf{$N\!aF\,$}

\begin{document}
\normalsize
\par
\begin{flushright}
ISN report 01-89 \\
\today
\end{flushright}
\par
\vspace{1cm}
\begin{center}
{\Large \bf Experimental study of a proximity focusing \cer counter prototype 
for the AMS experiment}
\vspace{0.3cm}
\par
\large{T.~Thuillier, F.~Malek, G. Boudoul, J.~Ballon, A.~Barrau, J.~Berger, M.~Bu\'enerd
\footnote{Corresponding author: buenerd@in2p3.fr}, L.~Gallin-Martel, 
A.~Menchaca-Rocha\footnote{Permanent address: Instituto de Fisica, IFUNAM, Ap. Postal 
20-360, Mexico DF}, and J.~Pouxe}
\par
{\sl Institut des Sciences Nucl\'{e}aires, IN2P3, 
53 av. des Martyrs, 38026 Grenoble cedex, France}
\par
\normalsize
\end{center}
\vspace{0.5cm}
\begin{center}
\parbox{16cm}{\underline{abstract}: A prototype of Proximity Focussing Ring Imaging \cer 
counter has been built and tested with several radiator materials using separately 
cosmic-ray particles and $^{12}$C beam fragmentation products at several energies. Counter 
prototype and experimental setup are described, and the results of measurements  
reported and compared with simulation results. The performances are discussed in the 
perspective of the final counter design.
}
\end{center}
\setcounter{page}{1}
\section{INTRODUCTION}\label{INTRO}

Proximity Focussing Ring Imaging counters (PFRICH) are based on a very simple geometrical 
configuration. The counter principle consists of a simple thin solid or liquid radiator, 
separated from the photodetector plane by a gap allowing photon rings associated to \cer 
cones to expand and reach a suitable radius before they are detected (see \cite{FRICH,YPS} 
for a general overview of RICH counters). 
\par
The price to pay for this architectural simplicity is a modest velocity resolution of the 
counter with respect to the best achievable performances \cite{HYPO}. This type of 
configuration is suitable for counter designs requiriring a large geometrical acceptance 
\{detection area\}$\otimes$\{angular range\}, for which the use of focussing devices is 
severely limited \cite{FRICH} or even impracticable, provided the required velocity 
resolution is not too high. The limiting resolution of these counters is set by the 
chromatic dispersion of the radiator material. In practice, the thickness of the radiator 
used as well as the spatial resolution of the photodetector array are also limiting factors to the 
counter resolution. The issue has been extensively discussed in a previous report on a 
simulation study of the counter \cite{SIMU} which complements the present work. Some of the 
results of this study will be repeated here for the reader's convenience.
\par
The AMS project consists of a particle spectrometer scheduled to be installed aboard the 
International Space Station (ISS) by the year 2004 for a 3 to 5 years campaign of 
measurements, with a broad physics program \cite{AMS,BE10}. The spectrometer will include 
a RICH counter among its instruments. The purpose of this counter is to achieve particle 
identification with the resolution performances shown to be realistic in the simulation study
for mass and charge measurements. These are: \\
a) A one $amu$ (atomic mass unit) mass separation for light nuclei over a broad momentum range 
extending from about 1~GeV/c per nucleon, up to around 13~GeV/c per nucleon  at best, for 
mass numbers A$\approx$20. This could be obtained by combining two radiators as shown in 
\cite{SIMU}. \\
b) A one charge unit separation for nuclei up to $Z\approx$~25 at best, for charge measurements, 
over the full momentum range of the spectrometer, i.e., from threshold up to above 1~TeV per 
nucleon. This latter performance depends critically on the electronics and PMT gain stability 
and calibration.  
\par
The full geometrical acceptance (${\cal S}\cdot\Omega$, ${\cal S}$ area, $\Omega$ angular 
acceptance) of the spectrometer will be of the order of $\approx$0.5~$m^2\cdot sr$ for the 
RICH $\otimes$ TOF $\otimes$ TRACKER combination of detectors. The overall spectrometer 
dimensions are restricted by rigid constraints on the payload envelope which must fit inside 
the space shuttle bay. These requirements were pointing to a PFRICH type solution because of
its simplicity, although alternative more ambitious options could have been taken. 
\par
The counter described here was a study prototype of first generation, built to perform an 
end-to-end test of the technique, from implementation of each component involved, up to 
velocity measurement, including the (first generation) prototype of front-end electronics, 
and event reconstruction algorithm. The purpose was to get through all the steps of the 
experimental procedure and to uncover all unexpected difficulties in order to finally 
reach the stage of the final counter design with a proven technique. 
The main points were: 1) investigating velocity and charge resolution capabilities of the 
counter over the full range of acceptance, in particular for large particle trajectory 
angles, 2) testing of the event reconstruction procedure, investigating potential background 
problems and their impact on the counter performances, and 3) testing the readout electronics. 
The prototype has been operated with cosmic-ray particles ($CR$) for several months, and 
tested with $^{12}$C ion beams at various energies at the GSI/Darmstadt facility.
\par
This article reports on the results obtained. The counter and its 
instrumental environment are described in section~\ref{APPAR}. The readout electronics are 
presented in section~\ref{DAQ}, with the data acquisition system used ($DAQ$), the latter 
for completeness. The analysis procedure is developed in section~\ref{ANA}, and the results 
are given and compared with simulation in section~\ref{RESUL}. The work is summarized and 
concluded in section~\ref{DISC}. 
\par
Some partial results of this work have been reported previously in a few contributions to 
conferences \cite{ANTEPROT}.
\par
\section{Description of the apparatus}\label{APPAR}

\begin{figure}[!hhh]
 \begin{center}
 \includegraphics[scale=0.3,angle=0]{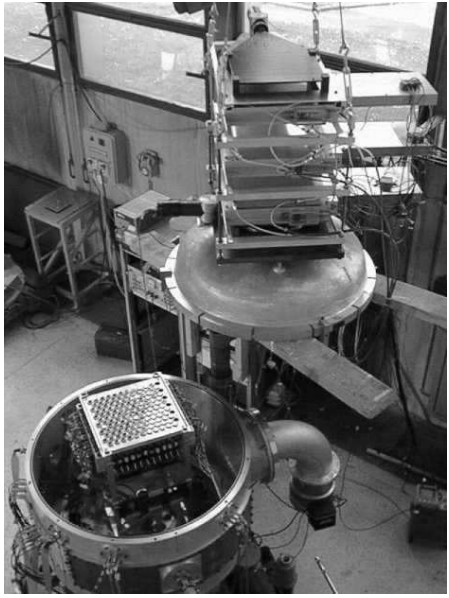}
 \end{center}
 \caption {Photographic view of the experimental setup during cosmic ray tests, showing the 
PMT matrix placed inside the vacuum chamber and the tracker (mwpcs) and trigger (scintillator 
paddles) system installed on the chamber lid, the latter being moved with a crane. }
 \label{PHOTO}
\end{figure}

The prototype consisted of a matrix of 132 3/4" diameter Philips XP2802 photomultiplier 
tubes (PMT) available from a previous experiment \cite{HYPO}. The size was compatible with 
the requirements defined from the preliminary simulation results. The PMTs were equipped with 
a lime glass window allowing photon detection over the wave length range \{280-640\}~nm. The 
tubes were mounted mechanically with individual magnetic shieldings on a support frame of 
aluminium drilled with appropriately spaced housing holes. Each PMT was mounted on a socket 
connected by a short cable to the front end electronics board adjacent to the matrix. The 
counter was installed in a vacuum chamber equipped with a pumping system for tests in 
vacuum. Two experimental configurations were used for cosmic ray and beam particle 
detection respectively. 
\par
In the two experimental setups, the prototype was complemented with a set of detectors
implemented to define the event, provide a trigger to the DAQ system, and allow the incident 
particle trajectory reconstruction.
\par
In the cosmic ray configuration, the counter surface was placed horizontally, facing the sky. 
A simple tracker made of 3 $xy$ multiwire proportionnal chambers (MWPC) 40x40~cm$^2$ with 2~mm 
wire spacing, and equipped with delay line readouts, was placed above the vacuum chamber (see 
fig. ~\ref{PHOTO}). It was used for incident trajectory reconstruction using the three space 
points provided by the MWPCs. The spatial resolution obtained for the extrapolated trajectory 
impact on the detection plane was about 1~mm in both directions, a value which did not affect 
sensitively the accuracy of the \cer event reconstruction. Three plastic scintillator paddles 
of different sizes read by PMTs, defining the angle of acceptance on the radiator, were 
interleaved with the MWPCs and used to define the trigger. They also provided dE/dX and time 
of flight (TOF) informations. 
Figure~\ref{PHOTO} shows a photographic view of the setup in cosmic ray configuration. 
\par
The beam test configuration is described in section \ref{BEAMTST}. Details on the setup and 
on the calibration procedures are given in ref~\cite{THESETH}.  
%
\par{\bf Radiators:}
Two types of radiator materials considered as suitable for the final counter have been evaluated, 
as discussed in ref \cite{SIMU}. \\
1) Sodium Fluoride (\naf), a crystal with low refractive index \cite{CA94}. It was 
chosen because of its suitability for low momentum particle identification (range 
$\approx$0.5-4 GeV kinetic energy per nucleon). \\
2) Silica aerogel (AGL). Several values of the refraction index were investigated because of 
their suitability for the intermediate and high momentum range of particle identification. 
One of them was used in the threshold \cer counter ($ATC$) which flew with AMS on the STS 91 
shuttle flight \cite{ATC}. The size and basic properties of the radiators for \cer 
light emission (mean refraction index $<\!n\!>$ , threshold velocity $\beta_{th}$ and momentum 
per nucleon (nuclei) $P_{th}$, limiting \cer angle $\theta_c^\infty$, photon yield, and chromatic 
dispersion) are given in table~\ref{DIMS}. The numbers are calculated taking into account the
\cer spectral distribution and PMT overall quantum efficiency. $<\!n\!>$ is the mean 
refractive index of the material used as radiator, $\beta_c$($P_{th}$) is the \cer velocity
(momentum) threshold. The momentum range $P_{range}$ is defined between the \cer emission 
threshold and the upper momentum limit defined by 4$\sigma$ separation of one $amu$ mass 
difference for a 1 cm thick radiator at the chromatic limit (see \cite{SIMU}). 
$\theta_c^{\infty}$ is the limiting \cer angle and $<\!N_{pe}\!>$ is the expected number of 
photoelectrons to be detected for $Z=1$ particle assuming that the full detection area is 
sensitive.
The experimental results for these radiators are available in Table~\ref{tab:syntheserun}

Both NaF and aerogels are transparent in the useful wave length range considered, extending 
from the upper UV region (300~nm) up to the (low yield) red region. See discussion below 
for the aerogels. The two materials combine very conveniently for maximizing the momentum 
range of particle identification \cite{SIMU}.
\par
\begin{table}
\begin{center}
\caption{\small\it Physical parameters of the radiator materials used in the study as 
described in the text.}
\begin{tabular}{lllllllll}
\hline\hline
Material  &$<\!n\!>$	&  Size   &  Thickness  & $\beta_c$ &   $P_{range}$      & $\theta_c^{\infty}$   & $<\!N_{pe}\!>$ & $\frac{\delta n}{n}$\\
          &		  &   (mm)     & (cm)        &             &     GeV/c/uma    &  (mrad)                     & (cm$^{-1})$      & $\times 10^3$   \\
\hline 
 NaF      &   1.33    	&  8.5$\times$8.5   &    1    & 0.75 & 	1-6.5	& 719 & 28 & $\sim 3$\\
 NaF      &   1.33    	&  8.5$\times$8.5   &    0.5   & 0.75 &  	& 719 & 28 & $\sim 3$\\
\hline 
 aerogel  &   1.14    &   4.1$\times$4.1  &    0.65  & 0.877 & 1.8-8	& 501 & 20& $\sim 2$\\ 
 aerogel  &   1.05    &   5$\times$5      &    2.5   & 0.952 & 2.9-10	& 310 & - & - \\
 aerogel  &   1.035   &  11$\times$11     &    1.1   & 0.966 & 3.4-11.5	& 261 & 6& $\sim 0.5$\\
 aerogel  &   1.025   &  11$\times$11     &    1.1   & 0.976 & 4.2-12	& 221 & 4 & $\sim 0.3$\\
 \hline\hline
\end{tabular}
\end{center}
\label{DIMS}
\end{table}
%
%
\section{Readout electronics and data acquisition}\label{DAQ}
%
\begin{figure}[ttt]
\begin{center}
\includegraphics[scale=0.45,angle=0]{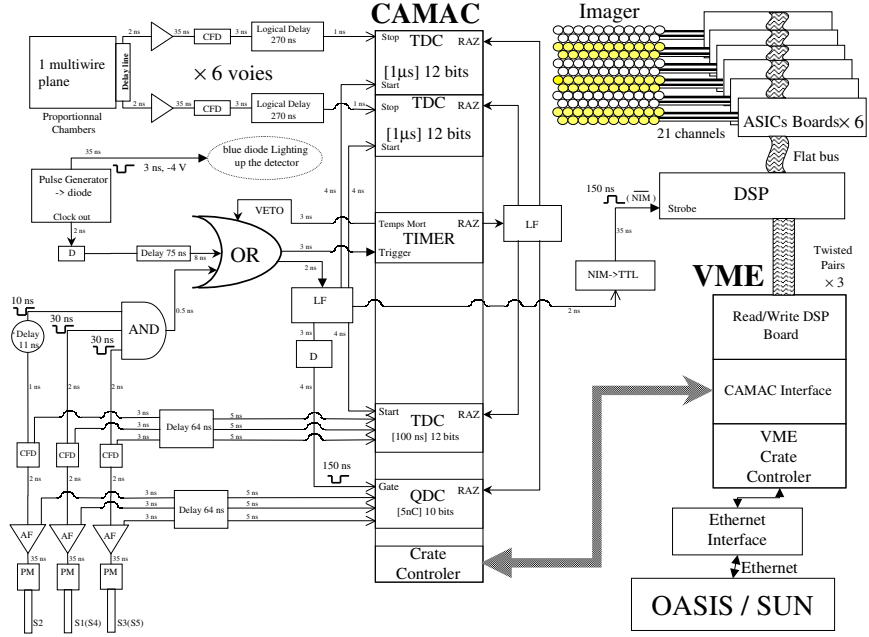}
\scriptsize{
\caption{\em 
Architecture of the signal processing electronics and data acquisition system used with the 
prototype.  The abreviations stand for:  
  CFD : Constant Fraction Discriminator,
  AF : Analog Fan-out, 
  LF : Logic Fan-out, 
  D : Amplitude Discriminateur, 
  TDC : Time to digital Converter, 
  QDC : Charge to digital Converter,  
\label{ELECTR}}}
\end{center}
\end{figure}

The design of the dedicated low power consumption readout electronics developed 
for this project is described in ref~\cite{GA99}. The readout and packaging system is built 
in a modular form composed of eight 32 channels processing boards connected to a bus 
together with a DSP board and a VME interface (dual access memory) for data storage and 
communication with the data acquisition system. The set up can process 256 
channels. In the present case, only 144 channels (6 boards of 24 channels) were used. See 
ref~\cite{NEWELEC} for the prototype II version currently in test phase (not used here). 
\par
A general layout of the electronics setup used in the measurements is shown 
on fig~\ref{ELECTR}. The trigger to DAQ was obtained by requiring a coincidence between the 
plastic scintillator paddles.
\par
The data were recorded by means of a general purpose data acquisition system ($DAQ$) 
allowing the online monitoring of the experiment~\cite{DAQ}. Part of the software has been 
developed by the authors for the purpose of the present study. The data were put on disk 
and transferred to a data storage facility. 
\section{Method of analysis}\label{ANA}
The analysis procedure was built along the following steps. First the geometrical alignments 
and calibrations required for the various detector were performed as described below. Next, 
each particle trajectory was first recontructed and then extrapolated onto the photodetector 
plane, providing the reference point for the Cherenkov pattern reconstruction. The 
validation cuts were then applied to the data, and for each photon of each event 
the \cer angle $\theta_c$ and the azimuthal angle $\phi_c$ were reconstructed individually 
using the algorithm described in \cite{SIMU}. Next, background photons were 
removed from the $\theta_c$ distribution, and finally the velocity of the particle was 
calculated using a weighted circular regression fit to the selected pattern, also described 
below.
\par
The charge Z of the particle was calculated in a separate step, from the total number of 
\cer photoelectrons measured for the event, by summing the response of the (calibrated) 
fired PMTs, using a dedicated background rejection procedure, and correcting for: a) The 
loss of internally reflected photons (NaF radiator only) according to the trajectory angle 
(see \cite{SIMU}), and b) The loss of refracted photons escaping laterally from the drift 
space of the counter. 
\subsection{Detector geometrical alignment}\label{DETAL}
For each run, the reference frame of the tracking system and of the PMT matrix had to be 
carefully aligned with respect to the photodetector reference frame. Indeed, a misalignment 
of the two detectors along transverse coordinates, generated a periodic dependence of 
the reconstructed \cer angle $\theta_c$ on the azimuthal photon angle $\phi_c$ measured with 
respect to the projection of the mid-radiator point of the trajectory on the detector plane, 
and a subsequent double peaking of the reconstructed $\theta_c$ distribution. 
A $\phi_c$-independent $\theta_c$ distribution can be easily obtained from a small sample of 
raw events by adjusting the $\Delta X$ and $\Delta Y$ transverse offset of the two 
coordinate systems. A good enough value of the $\Delta X$ and $\Delta Y$ offsets can be 
calculated using events with particle interacting with a PMT entrance window on the 
trajectory (see section~\ref{CUTS} for signal characterization). Assuming the PMT $X_{PM}$ 
and $Y_{PM}$ known geometrical coordinates on the detector plane, one can evaluate 
straightforwardly:
$$\left\{
\begin{array}{l}
\Delta X=\overline{X_{PM}-X_{track}}\\
   \\
\Delta Y=\overline{Y_{PM}-Y_{track}}
\end{array}
\right.
$$  
from the overall sample of data. The calculation of the mean is restricted to $|\Delta X|$ 
and $|\Delta Y|$ smaller than the PMT pitch ($\approx$25~mm). The calculation may require 
several iterations if the misalignment happens to be of the order of the detector pitch. 
\subsection{PMT gain and threshold alignments}\label{QGAIN}

The PMT sockets were grouped by 4 on a 40 channels High Voltage power supply. 
The mean gain of the PMTs on the detector was $G = 3.4\times 10^6$. On each processing 
board, the collected charge measurement was performed by a dedicated self-triggering analog ASIC 
micro-circuit ensuring charge to voltage conversion. The output voltage was digitized 
by a 12 bits analog-to-digital converter (ADC). A dynamic range of 100 has been chosen, 
based on estimate of the charge measurement range to be covered. This allowed the 
single photoelectron amplitude (SPE) to be encoded on about 40 channels. The electronic 
pedestal measurement of each channel was performed during dedicated runs. The PMTs 
calibration was performed using a blue LED diode. The diode light intensity was first set 
to a value and then reduced to reach the SPE signal on each channel of the detector. 
The calibration was then made assuming an exponential shape for the thermal (dynode) noise 
and a gaussian SPE response. In some cases, when the PMT response was poor, the second and 
third photoelectron peaks were taken into account \cite{BE94}. The gain measurement accuracy 
is estimated to be $\pm$5\%. The mean SPE resolution of the tubes was (RMS over mean value):
$\frac{\sigma}{q}\approx\, 50 \pm 14\,\%$, a reasonable result considering that
the PMTs used were ageing. 
\par 
The triggering level of each channel was set to approximately 0.3 SPE signal, so as not to 
lose SPE hits and to limit (thermal) background hits. It could be recorded directly in 
the ASIC memory during dedicated runs, the value being $\sim 15-20 \,mV$ under $50 \,
\Omega$, depending on the channel.
\subsection{Photon background}\label{BCKGD}
The photon background on the imager had to be treated with particular care since for $Z=1$ 
particles, the yield is small with $\sim$1-10 hit pixels per event, depending on the 
radiator material, and non-discriminated noise hits could damage considerably the accuracy 
on the velocity measurement. The mean overall background on the detector has been estimated 
from the data analysis to be of the order of 1-2 hits per event. Background hits triggering 
the ASICs can occur from the following sources:
\begin{itemize}
\item{PMT dark current: This contribution has been investigated by means of a random trigger 
generator during dedicated runs. The mean dark current (frequency of triggering) per PMT was 
relatively high with $f\approx 2500$ Hz because the PMTs were enclosed in a metallic black 
box in which the equilibrium temperature was elevated due to the PMT socket heat dissipation. 
As the acquisition time of the ASIC is $\tau_{asic} \approx 500$ ns, the probability to have 
one noise hit on the imager per event due to PMT dark current, could be estimated:
$$T_{dc}= 126\times f \times \tau_{asic} \approx \, 15\%$$ 
}
\item{Particle interactions in the PMTs can influence contiguous tubes. Cross-talk effects 
have been observed by studying samples of events that do not cross the radiator. The 
probability to have at least one fired PMT on the trajectory was $\approx 75\%$.
} 
\item{Reflected \cer Stray photons or Rayleigh scattered photons (for AGLs), reflected 
towards the detection plane. These photons were lost for the velocity measurement, but could 
be counted for $Z$ measurement. This contribution has not been investigated in details.
}
\end{itemize}

\section{Data selection}\label{CUTS}

Good events were selected by application of a set of software cuts to the data sample. 
A good event was defined by requiring valid particle trajectory, particle geometry, and 
\cer pattern.
\par\noindent
\begin{itemize}
\item{Valid particle trajectory:
Three space points from the three MWPCs and a chisquare $\chi^2<3$ in the linear regression 
on the 3 sets of $xy$ hit coordinates of the particle trajectory.
}
\item{{Valid particle geometry:} The extrapolated particle trajectory on the detector plane 
must cross the radiator.
}
\item{{Valid \cer pattern:}\\ 
At least three valid \cer photon hits, i.e., below the upper limit in amplitude, which
depends on the particule charge. The cut on the \cer pattern was processed in two 
steps. First, large amplitude hits lying around the extrapolated particle trajectory 
originating either from the \cer yield of particles crossing the PMT entrance windows, or 
from electrons produced in the crossing of dynodes, were excluded (referred to as the 
TC cut in the following). It consists of excluding the PMT crossed by the 
reconstructed trajectory and all the adjacent PMTs ( 7 pixel cut). The second algorithm was 
applied next to the remaining hit pixels for background hit rejection.
}
\end{itemize}
\par Two methods for \cer cluster identification in the $\theta_c$ angle distribution of 
hits have 
been tested. Both were applied after a preliminary sorting of the individual reconstructed 
\cer angles $\theta_{ci}$ by increasing value order. In the first algorithm, a cluster is 
identified as a set of \cer angles in which the angular distance between any two hit angles 
is smaller than a fixed value $\delta \theta_1$. A clustering weight $w_{i}$ proportional 
to the number of contiguous hits is associated to each $\theta_{ci}$ angle. The value of the 
weight of pixel $i$ then informs on its number of adjacent neighbours. For instance, only 
two adjacent pixels would give a 1-1 sequence in term of cluster weight, while a group of 3 
pixels would give 1-2-1. The raw hit multiplicity $\mu$ of an event is separated into two 
parts so that $\mu=s+n$, where $s$ is the number of hits assigned by the cluster algorithm 
while  $n$ is the sum of the rejected hits. In our study, a \cer cluster is kept if $s \ge 
3$ with $s > n$. Hence, a valid cluster contains at least one weight equals to two, which 
is an easy criterion to check, and the minimal weight pattern is 1-2-1.
In a normal run, events with several clusters were rejected. This algorithm has been 
successfully used to identify two independent \cer patterns in the double radiator 
configuration described in the section~\ref{TWORADS}. The $\delta \theta_1$ value has been 
investigated experimentally. It has been underlined that the $\delta \theta_1$ optimum is a 
compromise between $\chi^2$ likelihood (see section \ref{chi2}) and ring reconstruction 
efficiency $\epsilon$, defined as the ratio of the number of reconstructed rings 
on the number of events passing the cuts. These two numbers are calculated on the basis of the 
set of events with a validated reconstructed trajectory.
When $\delta \theta_1$ is too small, a fraction of good hits is rejected and $\epsilon$ 
decreases, while for $\delta \theta_c$ 
too large, $\epsilon$ is maximized but the likelihood is poor since noise hits have not been
efficiently rejected. For the prototype, the optimum value for NaF was $\delta \theta_1\approx 
35$ mrad while for 1.035 AGL radiator $\delta \theta_1\approx 80$ mrad. This method has given 
satisfactory results in rejecting noise hits far enough from the \cer cluster as it will be 
shown further below. 
\par  
The second method is simpler to compute and consists of assuming the median value 
$\theta_{cm}$ (of the sorted individual values of $\theta_c$) to be always located in the 
cluster. The valid \cer angles are then those for which $|\theta_c -\theta_{cm}| \le \delta 
\theta_2$. Here  the optimum value of $\delta\theta_2$ was around $50$ mrad for both Naf and 
AGL. The two methods give similar results.

\begin{figure}[ttt]
\begin{center}
\begin{tabular}{cc}
\includegraphics[scale=0.45,angle=0]{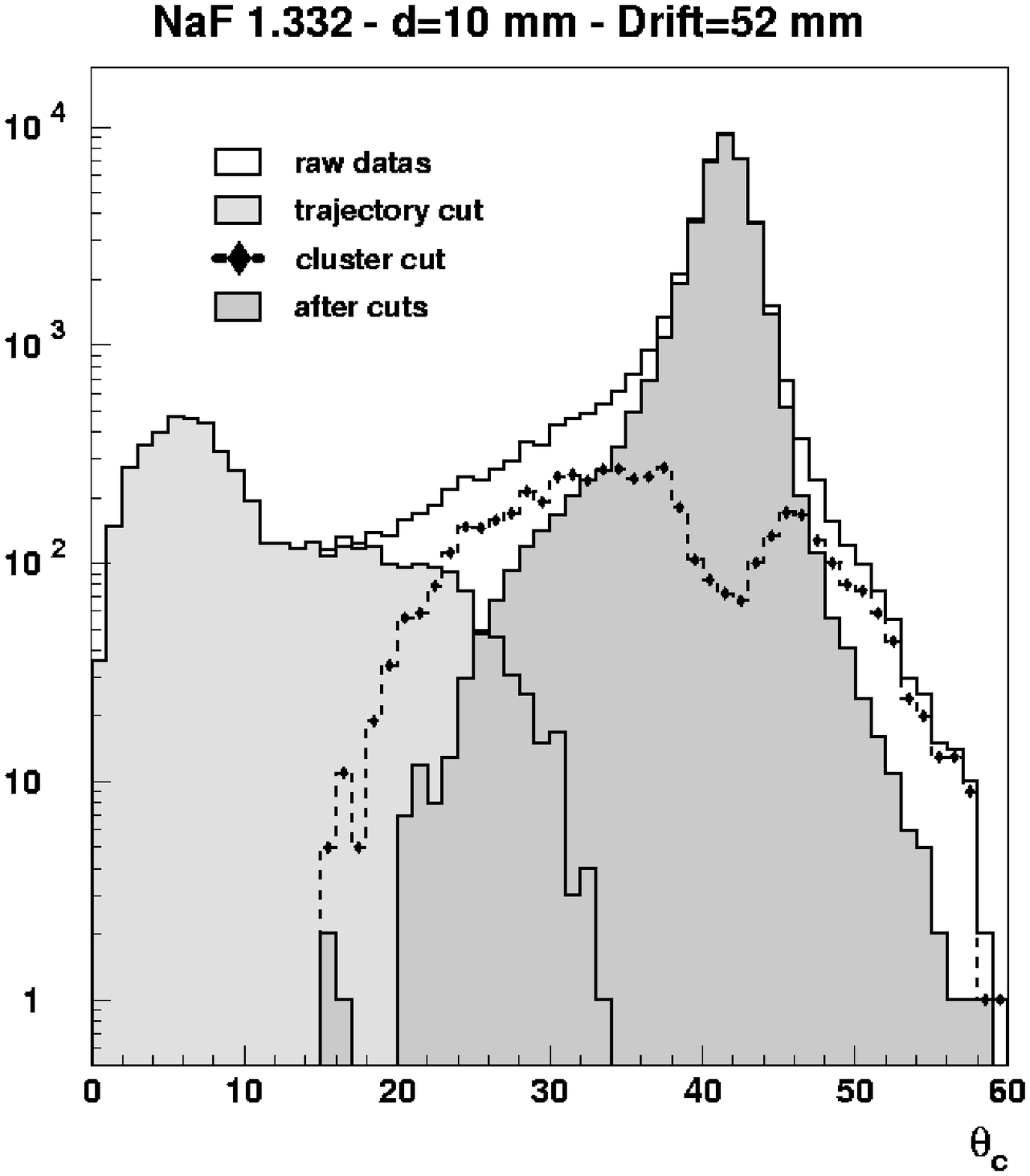} &  
 \includegraphics[scale=0.45,angle=0]{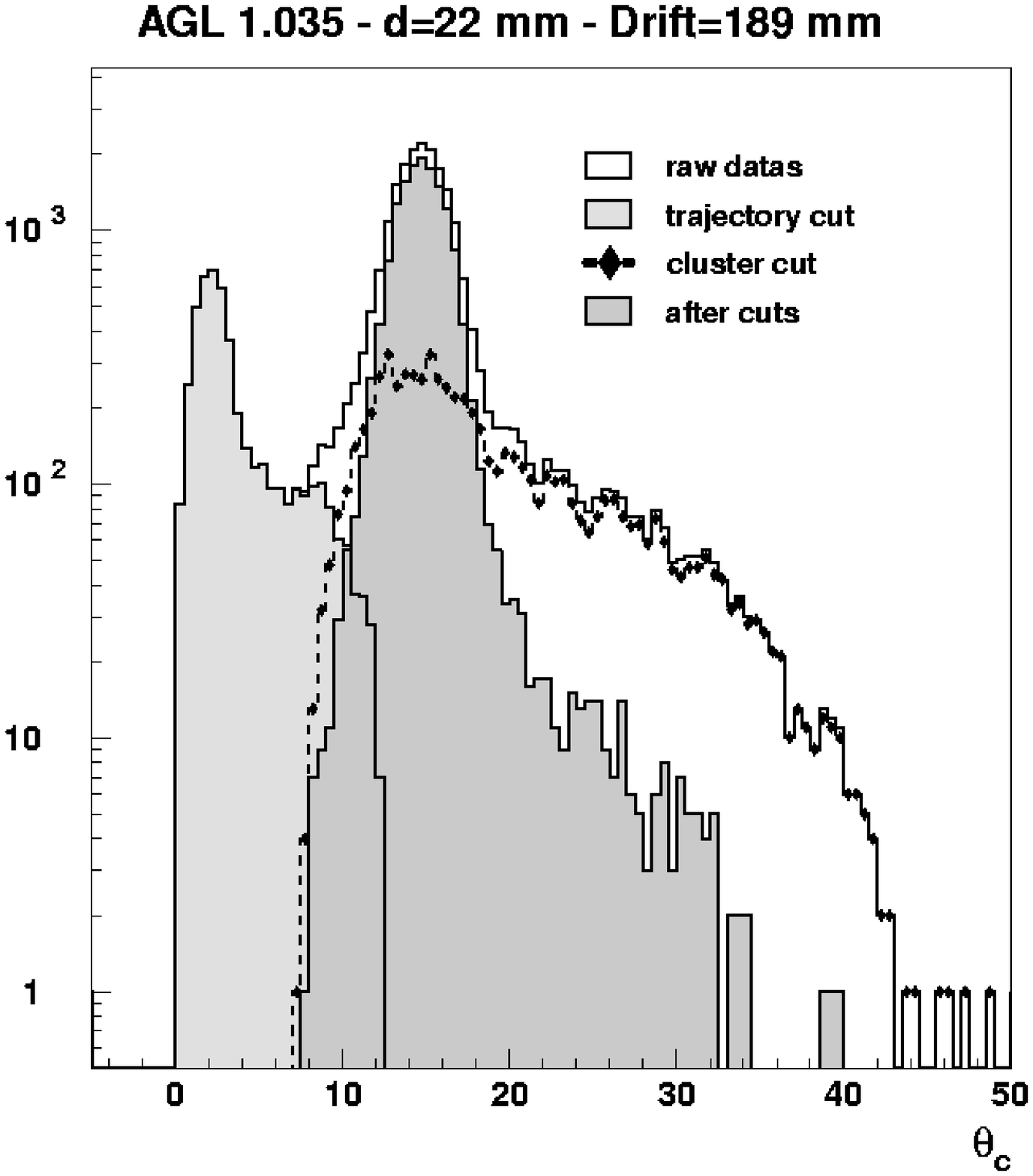}\\
 \end{tabular}
  \end{center}
  \caption{ \em Distributions of individual \cer angles $\theta_c$ obtained for:   
  10 mm NaF radiator with $52$ mm drift space (left), and 
  22 mm AGL radiator ($n=1.035$) with $189$ mm drift (right).
  Solid line : raw data; 
  Diamonds: individual $\theta_c$ cut by the cluster algorithm; 
  light gray: individual $\theta_c$ cut by the TC cut algorithm; 
  Dark Gray: remaining distribution of $\theta_c$ after both cuts have been applied. 
  \label{comparemethode}} 
\end{figure}

\begin{figure}[!hhh]
\begin{center}
\begin{tabular}{cc}
   \includegraphics[scale=0.4,angle=0]{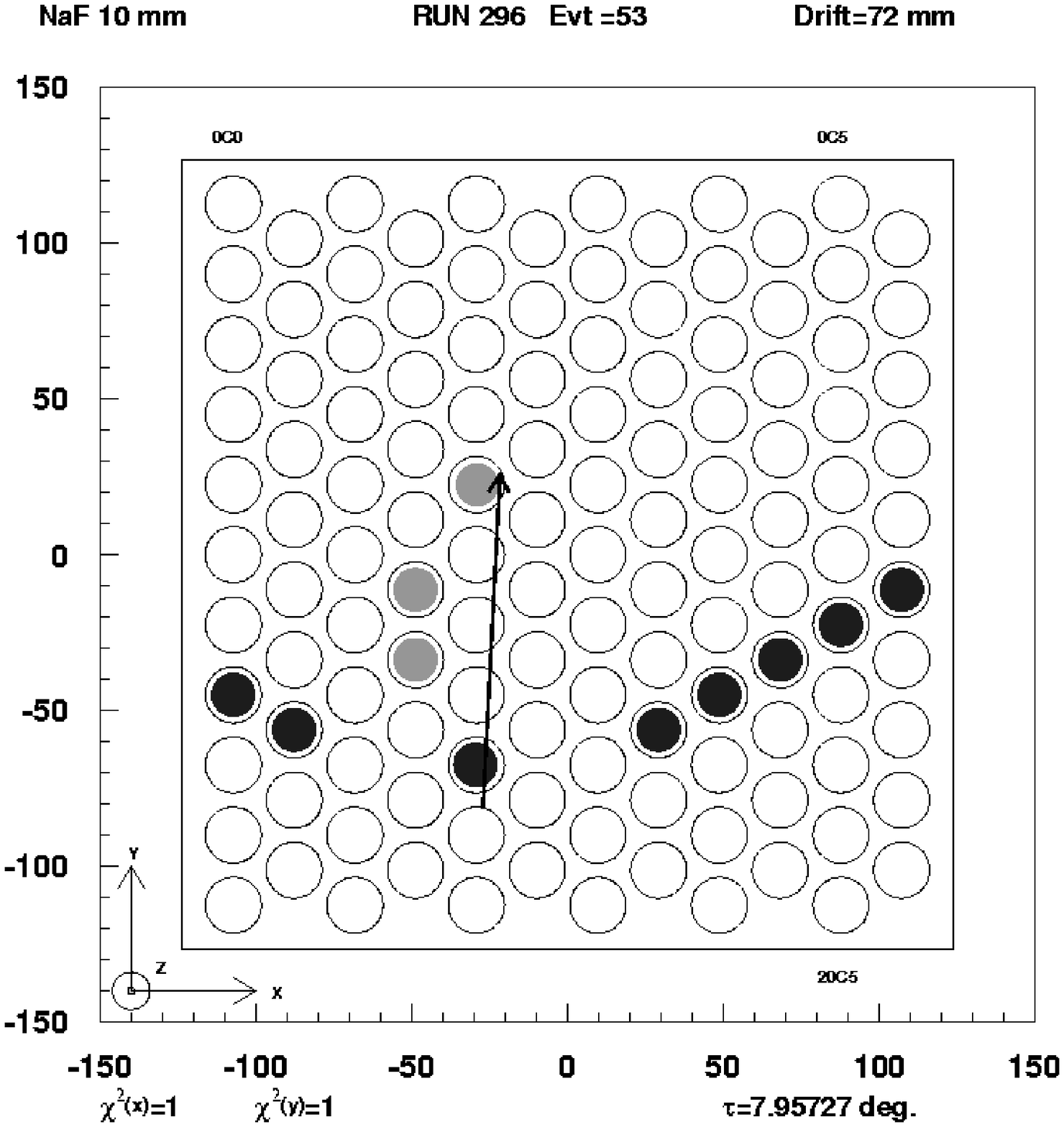}
   &
   \includegraphics[scale=0.4,angle=0]{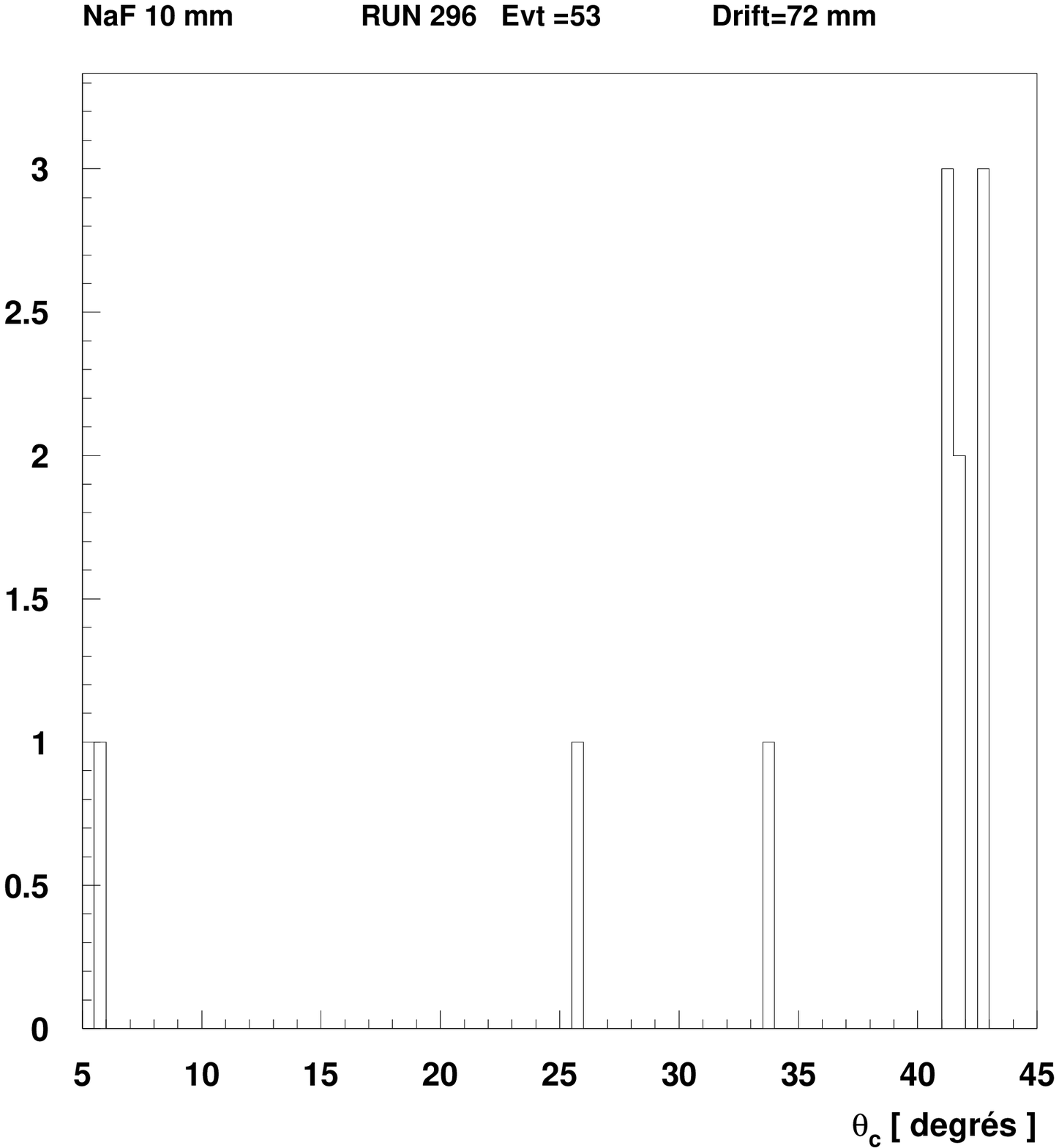}\\
   (a) & (b)\\
   \begin{minipage}[c]{7.5cm}
   \begin{centering}
   \vspace{-8cm}
  \caption{\em (a) Example of event obtained with a 10 mm thick NaF radiator and a 72 mm 
   drift gap. Full circles are \cer hits (black) and background hits (gray). The particle 
   was an atmospheric muon ($\beta\sim 1$) with too large a velocity to be resolved with the 
   prototype. 
   (b) Distribution of reconstructed $\theta_c$ values for this event hits, using the 
   particle track parameters. Three bakground hits are seen neatly separated from the \cer 
   cluster. 
   (c) Circular regression fit for the same event of the \cer ring in the trajectory frame 
   (rejected hits labelled *). \label{evtnaf}} 
   \end{centering}
   \end{minipage}&
   \includegraphics[scale=0.4,angle=0]{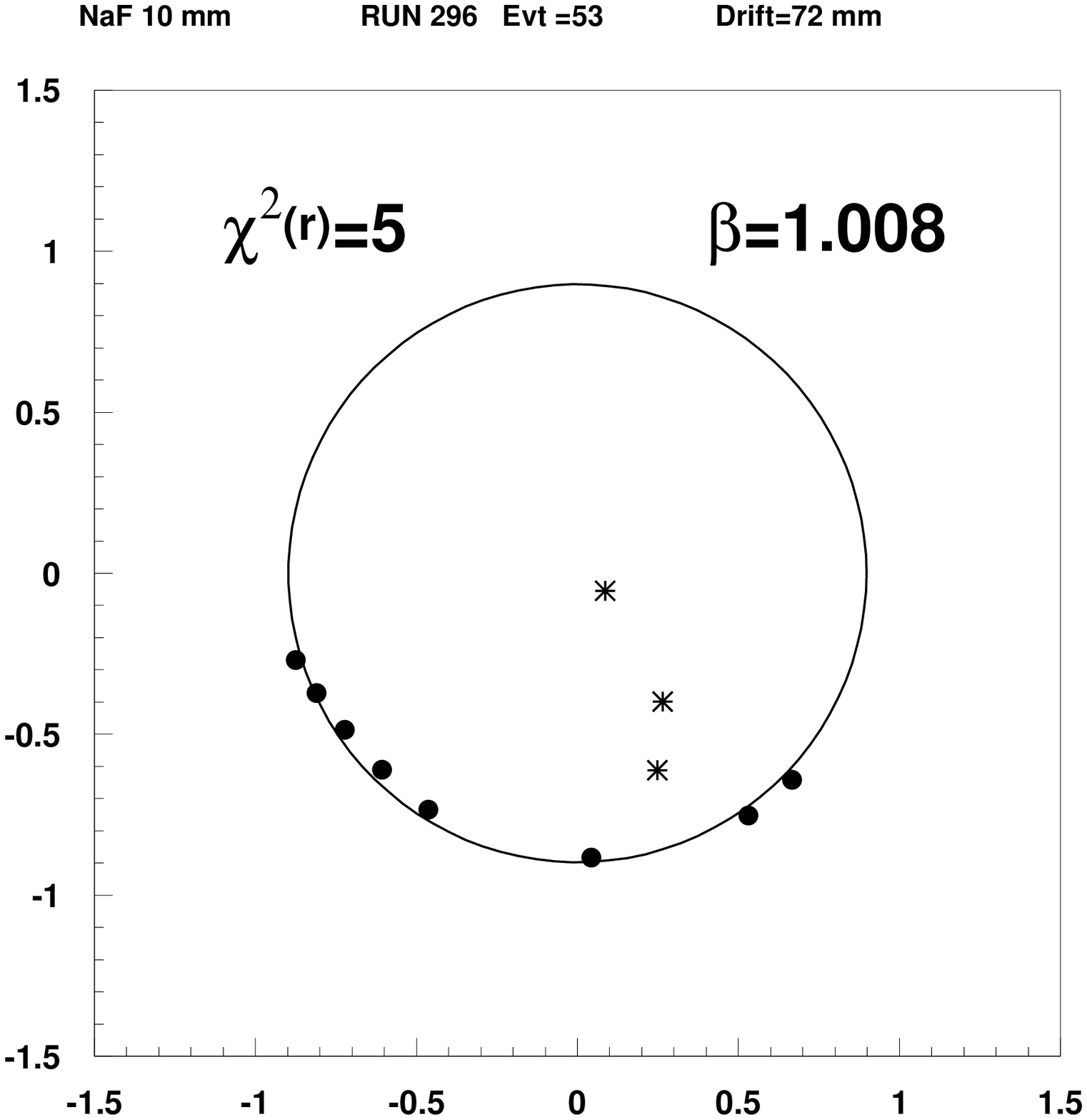}\\
    & (c) \\  
  \end{tabular}
  \end{center}
    \end{figure}

Figure~\ref{comparemethode} illustrates the results for two radiator materials. It shows the 
raw $\theta_c$ distribution (solid line histograms) for NaF radiator (left), and AGL 1.035 
radiator (right) for cosmic ray (CR) particles. On both figures, two peaks are seen at low and  
high $\theta_c$ values, the latter close to $\theta_c^\infty$ value, containing most of good 
events. The small angle peak is due to particles interacting with PMTs.
The specific cut of these low angle hits (TC cut) corresponds to the pale gray 
histogram. It consists of a peak at small angle riding on a broader distribution extending to 
higher $\theta_c$ values. The peak reflects the angle distribution between the center of the 
hit pixel and the impact point of the particle on the imager (the $\theta_c$ reconstruction 
algorithm takes the center of the PMT photocathode as photon hit coordinates \cite{SIMU}). 
On figure~\ref{comparemethode}(a), the peak is observed around 5~$^{\circ}$ while the broader 
structure extends up to 30$^{\circ}$. This structure is a consequence of the cross-talk 
effect between the hit PMT and its adjacent neighbours. Cutting these cross talk effects 
is necessary but has the unwanted drawback of cutting the \cer signal for particles having 
velocities close to the \cer threshold, since for small $\theta_c$ the reconstructed rings 
is contained within the first circle of PMTs around the particle impact. Thus, on the 
prototype, the effective $\beta$ threshold is somewhat higher than the $\frac{1}{n}$ value. 
This effect is expected to be limited however with smaller pixels as it will be in the final 
design. It will also depend on cross talk between pixels for the PMTs used.
\par 
On fig~\ref{comparemethode}, the part of the distribution cut by the cluster algorithm is 
represented by diamond histograms. The peaking around $\theta_c^\infty$ is seen to shrink 
significantly for both radiators, showing the efficiency of the cut. This is particularly 
clear for the AGL radiator where the high $\theta_c$ tail due to Rayleigh scattered photons 
is reduced by about one order of magnitude. Note that the cusp generated in the distribution 
for NaF is a small $\approx$1\% effect.
\par 
The distribution of $\theta_c$ remaining after cuts is displayed in dark gray. The large 
angle tail for AGL can be partly cut out by requiring individual $\theta_{ci}$ to be less than 
$\theta_c^\infty+3\delta\theta_1$ for instance. For Naf, the shoulder around $30^\circ$ is probably 
physical since the hit multiplicity observed for these events is consistently decreasing 
with $\theta_c$ as expected for \cer photons.
\par 
An example of CR event measured with NaF radiator is displayed on figure~\ref{evtnaf}(a), 
with the matrix of PMT array represented. The arrow shows the projection on 
the detector plane of the particle trajectory between the entry point in the upper MWPC and 
the intercept on the detector plane (arrow head). 
The reconstructed individual $\theta_c$'s for this event are shown on figure~\ref{evtnaf}(b).
The \cer cluster is easily identified to the peak on the right, close to $\theta_c^\infty$.
The noise hit at $\approx 6^{\circ}$ is removed by the TC cut. The other two 
background hits are rejected by the cluster algorithm.
\section{Particle velocity reconstruction}\label{chi2}

On the detection plane, the \cer pattern generated by refraction of the photons at the 
radiator exit interface is a complexe curve. A simple circular regression fit procedure can 
be used however by transforming the hit coordinates to the trajectory frame where the \cer 
light is uniformly distributed on a cone, and the hits along a circle. 
The individual azimuthal angles $\phi_c$ defined below in the detector frame, can be 
expressed to give new azimutal $\psi_c$ angles in the trajectory frame. The parametric 
equations of the circle are given by (per unit of length) :
$$ \left\{
	  \begin{array}{l}
	   x =\tan \theta_c \cos \psi_c \\
	   \\
	   y= \tan \theta_c \sin \psi_c\\ 
	  \end{array}
    \right.
$$

The free parameter of the fit appears to be the reconstructed radii: $r_i=\tan \theta_c{_i}$,
and the $\chi^2$ function becomes:
$$\chi^2=\sum_{i=1}^N \frac{\left(r_i^2-r^2\right)^2}{\sigma_{r^2_i}^{2}}$$
The uncertainty $\sigma_{r^2_i}$ on $r_i^2$ is calculated numerically pixel by pixel, for 
any incident particle angle. $\sigma_{r^2_i}$ includes contributions from the hit pixel 
multiplicity, its size and position according to the particle trajectory, radiator 
thickness, trajectory uncertainty, radiator chromatism, and multiple scattering of particle 
in the radiator. A dedicated analytical approach of these quantities has been developed  
in reference \cite{THESETH}. See also \cite{HYPO} for the circular regression technique.
The ring radius $r_c$ minimizing the $\chi^2$ function is then :
$$r_c^2=\frac{\sum_{i=1}^N \frac{r_{i}^2}{\sigma_{r^2_i}^{2}}}
{\sum_{i=1}^N \frac{1}{\sigma_{r^2_i}^{2}}}$$
from which the experimental velocity $\beta$ can be derived:
$$\beta=\frac{1}{n}\sqrt{1+r_c^2}$$
An example of reconstructed velocity with $\chi^2$ minimization is available on 
figure~\ref{evtnaf}(c). It is the last step of the event reconstruction process illustrated 
on figure~\ref{evtnaf}(a). 
\section{Measurements with cosmic ray particles}\label{RESUL}

The average count rate was of the order of 0.2~s$^{-1}$. A typical run duration was 
one to three days, providing on the average 17000 events per day. 
%
 \begin{table}
 \begin{center}
  \begin{tabular}{cccccccccc} 
 \hline
 \hline
  radiator &$<\!n\!>$ & $d$  & $D$ & $\epsilon$  & $\epsilon^*$ & $ \frac{\delta 
                                                     \beta}{\beta}$ & $N_{pe}$ & $N_{pix}$\\
            &       & [mm] & [mm]& [\%]        & [\%] & $\times 10^{3}$  & & \\
 \hline
   NaF      & 1.332 & 10   & 52  & $91\pm 4.1$ &$>91$ & 8.8 & 19.5&  7.7 \\
   AGL      & 1.14  & 13   & 110 & $60.8\pm2.3$&$>60.8$ & 9.0 &  7.2 &  4.2 \\
   AGL      & 1.05  & 25   & 220 & $>80\pm2.5$ &$>85.1$ & 4.7  & 8.2 & 5.8 \\
   AGL      & 1.035 & 22   & 189 & $58.1 \pm 2.5$ & 62.4 & 4.3  & 4.8 & 3.4 \\
   AGL      & 1.035 & 33   & 245 & $67.0 \pm 2.1$ & 71.3  &3.5  & 7.1 &4.7  \\
   AGL      & 1.025 & 23   & 321.8 & $51.8 \pm 1.7$ & 58.3 & 2.7  & 4.9 & 3.0 \\
   AGL      & 1.025 & 34.5 & 310.3 & $65.1 \pm 1.9$ & 73.2 &2.8 & 6.6 & 4.0 \\
   AGL      & 1.025 & 46   & 298.8 & $66.1\pm 2.2$& 74.4 & 2.7  &7.3 & 4.3 \\   
   \hline
   \hline
   \end{tabular}
  \caption{\em List of radiators studied in CR tests. Here, $<\!n\!>$ is the mean refraction 
index of the radiator over the spectral range of the PMTs for \cer light, $d$ is the 
radiator thickness, $D$ the drift gap, $\epsilon$ the ratio of the reconstructed event 
number over trigger number, $\epsilon^*$ the reconstruction efficiency corrected from the 
cosmic ray spectrum below \cer threshold, i.e., triggering scintillators but producing no 
\cer light for particles down to $\beta=0.88$~\cite{cosmicray}. 
$\frac{\delta \beta}{\beta}$ is the velocity resolution obtained from a gaussian fit around 
$\beta=1$. $N_{pe}$ and $N_{pix}$ are the mean number of photons and the mean number of 
firing pixels per event after cuts, respectively.
     \label{tab:syntheserun}}
   \end{center} 
  \end{table} 

\subsection{NaF radiator}\label{NAFR}

The mean refraction index calculated taking into account the \cer light energy distribution 
and the quantum efficiency of the PMT photocathode is $<\!n\!> = 1.332$.
This material has been successfully used previously in the CAPRICE balloon experiment 
\cite{CAPR}. 
The maximum \cer angle is  $\theta_c^\infty\approx 41^{\circ}$. The refraction outside the 
radiator increases this angle to $\approx 61^{\circ}$ for particles normal to the detector 
plane. In practice, this value limited the drift distance usable on the prototype to a small 
range around $4-5$ cm for the full ring to be contained in the detector surface.

A good reconstruction efficiency of about $90\%$ was obtained for CRs with this radiator 
because of its 
high light yield and good transparency (see table \ref{tab:syntheserun}). The mean 
multiplicity was $7.7$ after cuts for 1~cm thick radiator. The best velocity resolution 
achieved with the NaF was $\frac{\delta \beta}{\beta}=8.8\times10^{-3}$. This rather poor 
value is mainly due to the small drift distance mentionned above which, combined with the 
large pixel size (1.8~cm) lead to a large uncertainty on the ring radius measurement. 
The large 
chromatic dispersion for this radiator lead to a contribution of the same magnitude as the 
former contributions to the overall uncertainty on the velocity measurement \cite{SIMU}. For 
large incidence angles, a significant fraction of the ring is internally reflected inside 
the radiator and is lost for detection. This effect however, is not expected to deteriorate 
the velocity resolution \cite{YPS,SIMU}.

\subsection{Aerogel radiators}\label{AGLR}

Several Silica Aerogel (AGL) samples have been tested in the prototype with the refraction 
index values $n=1.14, 1.05, 1.035$ and $1.025$. This type of radiator allows to fill part of the 
gap in the range of usable refraction index between gas and solid radiators. They became 
widely studied and used recently \cite{FI94,DE97,AS98,GO99,AS00,DE01,NA98} because of their 
low refraction index and chromatism compared to cristals (roughly a factor ten smaller), 
these quantities being correlated, see appendix and \cite{SIMU,VI01} for details. 
One drawback of AGLs is the Rayleigh scattering phenomenon due to the microscopic structure 
of the material. Photons are scattered inside the radiator material, and loose their angular 
coherence. The scattering cross section is larger for short wave lengths \cite{FI94}, and 
scattered photons generate a wide background halo around the unperturbed \cer ring. AGLs 
however are currently the only available material to cover certain range of refraction index 
and then of particle velocity  (see \cite{SIMU} for details).
The results on the velocity resolution and the experimental photoelectron yield for AGLs 
are summarized in table \ref{tab:syntheserun}, where it is seen that the velocity 
resolution globally improves with the decreasing refraction index (and correlated chromatic
dispersion), as it can be expected from general considerations \cite{SIMU,THESETH}. 
The velocity resolution can be expressed in terms of chromatism and of the uncertainty on 
the $\theta_c$ measurement per photon, from the \cer relation $\cos\theta_c=\frac{1}{\beta n}$ :
$$\frac{\delta\beta}{\beta}=\frac{\delta n}{n}+ \tan\theta_c\, \delta \theta_c$$
$\delta\theta_c$ being the experimental uncertainty on the $\theta_c$ measurement. 
It is clear from this well known relation that the velocity resolution improves with the 
decreasing chromatic dispersion provided the uncertainty on $\theta_c$ is kept smaller than 
the latter.
\par
A velocity resolution $\frac{\delta \beta}{\beta}=9\times10^{-3}$ could be obtained 
with the AGL 1.14 sample. Although the sample was apparently of rather poor optical quality, 
the results are close to the chromatic limit (see discussion in section \ref{DETAL}). 
The AGL 1.05 sample \cite{MATSU} provided a significantly better value, 
$\frac{\delta \beta}{\beta}=4.7\times10^{-3}$, and the best reconstruction efficiency of
the AGL sample tested, probably in account of the good transparency of this sample. 
The AGL 1.035 sample used \cite{MATSU} was part of the spare tiles from the Aerogel 
Threshold Counter built for the AMS01 experiment \cite{ATC}. The measurements provided a 
value of the resolution $\frac{\delta \beta}{\beta}=3.5\times10^{-3}$, with a reconstruction 
efficiency however, decreasing down to around $70$\% for a 3 cm radiator thickness, with 
respect to the AGL 1.05 sample, due to the lesser clarity of this material than for the 
1.05 index.
The AGL 1.025 sample \cite{MATSU} provided the best velocity resolution obtained in the tests, 
$\frac{\delta \beta}{\beta}=2.7\times10^{-3}$. Several runs made with increasing radiator 
thickness showed that above $\approx 3$~cm, both the reconstruction efficiency (for Z=1 
particles) and the velocity resolution remained approximately constant around $70$ \% and 
$2.7\times10^{-3}$ respectively. This was expected since, when the radiator thickness 
increases, the net gain of non-scattered \cer photons drops rapidly (+1.7 SPE between 2 and 
3~cm, and only +0.9 between 3 and 4).
\par
It can be observed in table \ref{tab:syntheserun} that for aerogels the achieved resolution 
scales with (n-1)/n as expected for the chromatic limit \cite{SIMU}. However the 
contributions to the resolution in all cases of the table are dominated by the pixel size 
contribution. The latter nevertheless follows closely the value of the chromatic 
contribution, generating the observed effect.
\subsection{Stability of the long term \cer light yield of silica aerogel}\label{AGIG}

\begin{figure}[!hhh]
 \begin{center}
 \includegraphics[scale=0.4,angle=0]{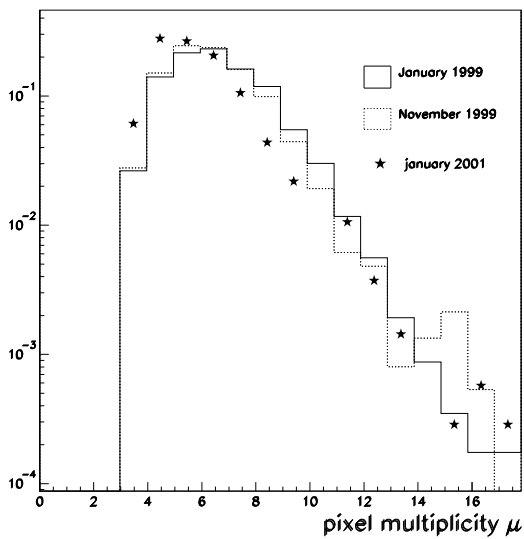}
 \end{center}
 \caption { Ratio of the hit multiplicity distribution to the number of triggers, 
measured for the three runs of january and november 1999, and january 2001, as 
discussed in the text. One run is not shown for the legibility of the figure.}
 \label{AGL_AG}
\end{figure}

The rapid decrease in time of the \cer yield of the AGL in the ATC counter observed in the 
AMS01
experiment \cite{ATC} has been a major concern for the AMS collaboration, the stability in 
time of the \cer response of AGL becoming an issue. It was pointed out in a previous note 
\cite{AGING} that the ATC AGL tiles have been processed and conditionned with chemically 
active products like solvants and wave length shifter, and that a chemical contamination was 
more likely than an (unoticed so far) ageing phenomenon of Silica. Although the observed 
effect is still awaiting for a proven explanation, the issue has been addressed 
experimentally with the study prototype and the \cer yield of the AGL monitored over about 
two years (January 1999 to january 2001). During this time, 4 runs have been recorded in 
identical conditions, using an AGL 1.035 sample from the AMS01 spares, by 6 months intervals 
of time, providing hit multiplicity distributions shown on figure \ref{AGL_AG}. 
Unfortunately, the prototype has been accidentally exposed for several hours to attenuated 
day light before the third run (june 2000) was taken, resulting in a significant decrease 
of the overall PMT detection efficiency by about 30\%. However, runs on NaF were measured 
after each AGL runs, and  the AGL/NaF ratio of the \cer yield was the same within a small 
statistical uncertainty before and after the accident, thereby providing a mean to normalize 
the late runs with respect to the early ones (table \ref{AGG}). 
As seen on figure~\ref{AGL_AG}, the results are unambiguous: No significant decrease of the 
mean multiplicity could be observed, and no evidence could be obtained for a natural ageing 
process of the material. The results are summarized in table \ref{AGG}. See \cite{AGING} for 
other details and a discussion of the origin of the decaying light yield observed for the ATC 
counter in AMS01. 
\begin{table}[ht]
\begin{center}
\caption{\small\it  Fraction of reconstructed events for the 4 runs measured in the same
conditions with the RICH study prototype and n=1.035 aerogel (two tiles 1~cm thick 
superimposed) over a two year period of time. The second and third columns give the ratio 
of reconstructed events (above threshold) over valid triggers (trajectories) and the 
associated statistical uncertainties in percent, respectively. Note that this is not the 
reconstruction efficiency since valid triggers include particles below the \cer threshold 
($\approx$4~GeV/c). The fourth column shows the $NaF/AGL$ counting ratio for the four runs,
while the last column gives the mean value of the $AGL$ hit multiplicity distributions 
shown on figure \ref{AGL_AG}. The yields for the last two runs have been normalized to the 
$NaF$ values. The uncorrected values are given in parenthesis (see text).}
\vspace{0.2cm}
\hspace{-1.5cm}
\begin{tabular}{lllll}
\hline 
  Date of data  & Reconst. evts  & Error    & Ratio   &  Mean hit     \\
                & \% of triggers & $\sigma$ & $NaF/AGL$ &  multiplicity \\
\hline \\
  Jan 1999      & 56.6           & 1.7     & 0.571    &   6.31     \\
  Nov 1999      & 55.5           & 2.3     & 0.556    &   6.55     \\
  Jul 2000      & 40.6           & 1.4     & 0.532    &   6.27 (5.72)   \\
  Jan 2001      & 43.4           & 1.6     & 0.543    &   6.41 (5.84)   \\
\hline 
\end{tabular}
\end{center}
\label{AGG}
\end{table}
\subsection{Dual radiator configuration}\label{TWORADS}

\begin{figure}[!hhh]
 \begin{center}
 \includegraphics[scale=0.5,angle=0]{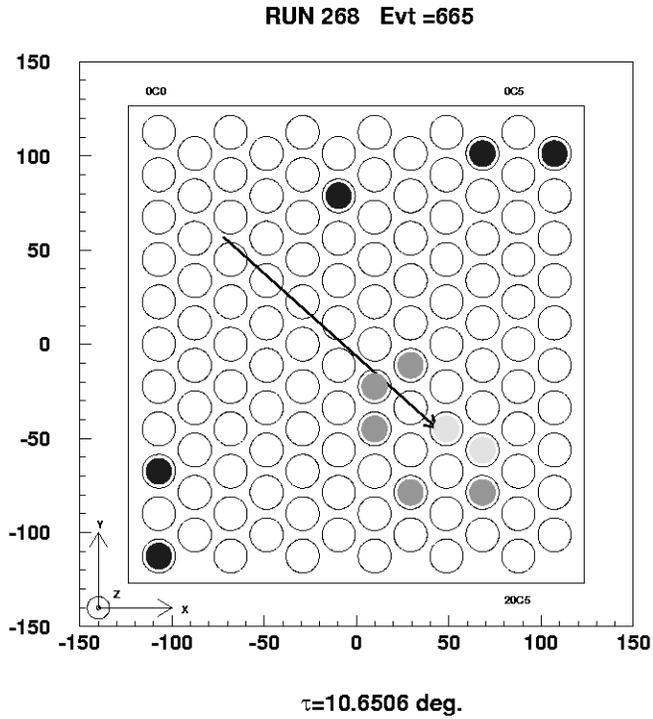}
 \end{center}
 \vspace{-1cm}
 \caption {\sl\small Example of double ring event obtained with the dual radiator setup. 
$\tau$ is the incident particle angle on the radiator. }
 \label{doubleevent}
\end{figure}

The interest of using a dual radiator system lies in the extension of the momentum range 
covered by the counter for a single radiator. In the present case, the idea was to combine 
NaF and AGL materials, in order to reach a broad identification range over the full counter 
fiducial area, allowing identification of ions from the NaF threshold at 480~Mev kinetic 
energy per nucleon, up to the upper limit for AGL 1.025, between 13 and 20 GeV/n for ions 
with mass about A=25 and A=4 atomic units respectively \cite{SIMU}. This broad range of 
sensitivity was dictated by considerations on the Physics case for the AMS RICH \cite{BE10}.
Since the \cer angles are very different for the two media, the risk of confusion is 
minimized, and the only difficulty to overcome was to find the appropriate way of 
processing the double hit-pattern.

\begin{figure}[!hhh]
 \begin{center}
 \includegraphics[scale=0.5,angle=0]{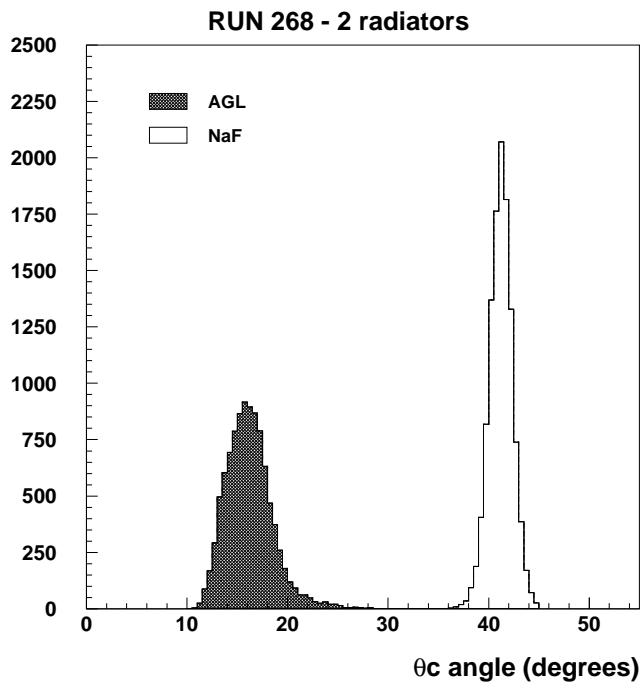}
 \end{center}
 \vspace{-1cm}
 \caption {\sl\small Distribution of the reconstructed \cer angles for the double radiator 
run, with AGL 1.035 (hatched histogram) and NaF (solid line). See text for details.}
 \label{doublethetac}
\end{figure}
\begin{figure}[!hhh]
 \begin{center}
 \includegraphics[scale=0.5,angle=0]{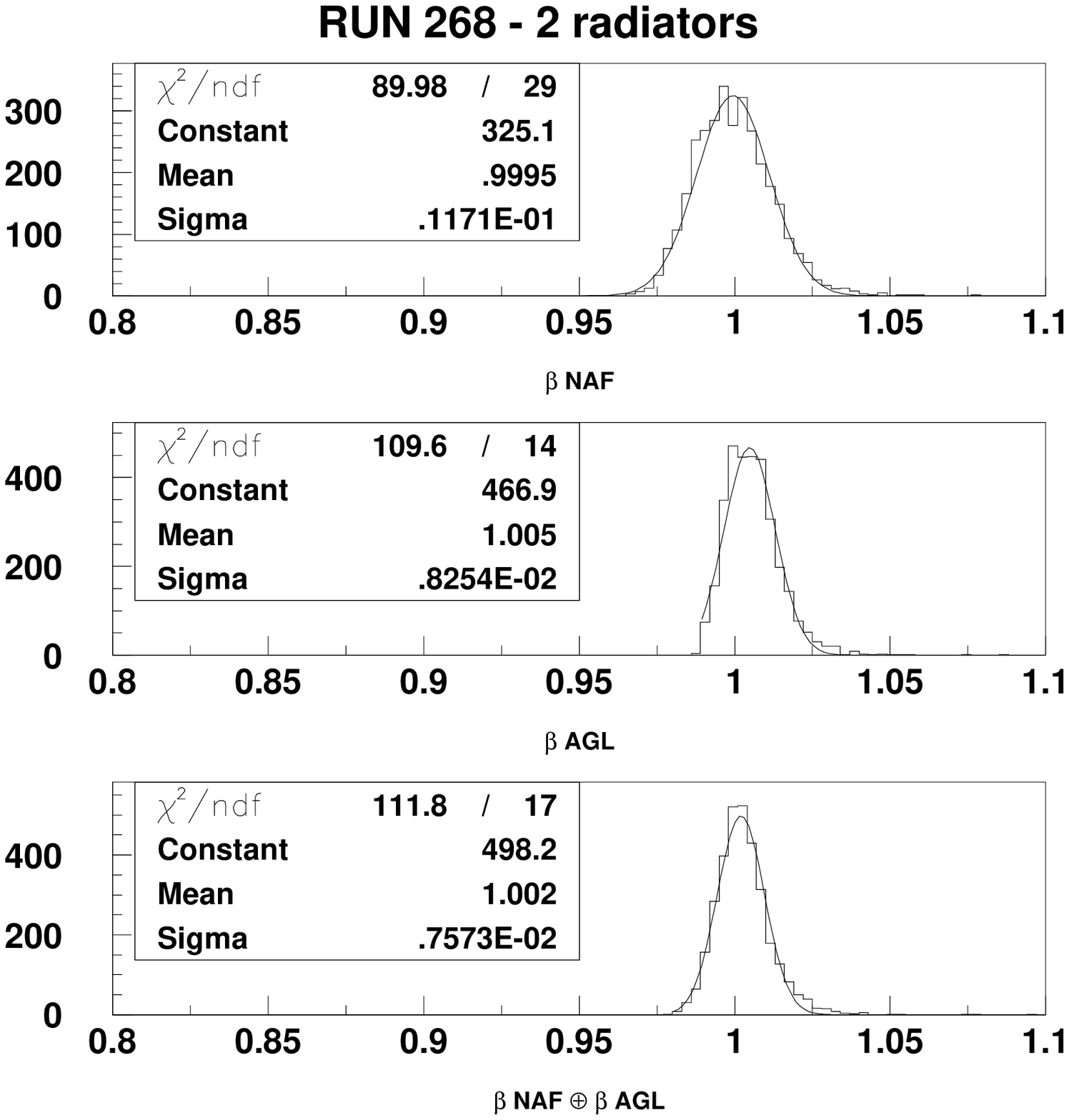}
 \end{center}
 \vspace{-1cm}
 \caption {\sl\small Double $\beta$ measurement obtained in the dual radiator run. From top 
 to bottom : $\beta$ measurement with NaF, AGL, and mean of the two measurements.}
 \label{doublebeta}
\end{figure}

\par 
A dedicated CR run was performed with the prototype, combining a 5 mm thick NaF 
and a 22 mm thick AGL (1.035) put together in a stack, to investigate the feasability of
reconstructing simultaneously the two \cer patterns. The AGL tile was placed above and the 
total drift distance between the NaF tile and the matrix was 110 mm. This configuration is
a compromise, rather far from the individual optimum geometry for the two radiators: for 
this drift distance, the ring size of the NaF photon pattern is of the 
order of the matrix size, while for the AGL photons it is of the order of the pixel size. 
Hence, a number of NaF photons is expected to be lost out of the detector, while the small 
AGL pattern is very closed to the lower limit of reconstruction imposed by the TC cuts
 (see section \ref{CUTS}). This problem however should vanish with the smaller pixel
size and large detector area of the final counter.

The double pattern reconstruction used in the analysis was based on the same cluster 
algorithm as developped in section \ref{CUTS}, with some modifications. 
First, the events were processed assuming all the photons to be produced in the upper (NaF) 
radiator. Next, The TC cut was applied, as for a normal single radiator event 
processing to reject particle-impact related low $\theta$ angles. The following step remained 
unchanged : the remaining angles were sorted and processed by the cluster algorithm. While in 
a normal single radiator run, events with more than one cluster were rejected, in the double 
radiator, double cluster structures were selected as valid events. Since the number of 
lighted pixels was small, the first type of double structure accepted was $1-1-1-1$ (since 
one single cluster including all hits would be $1-2-2-1$, see section \ref{CUTS}), which 
means only two pixels fired for NaF and two for AGL. The small $\theta_c$ cluster was 
assumed to be AGL. On this basis, the AGL $\theta_c$ angles were calculated taking into 
account the refraction in the NaF tile.
An example of double ring identification is displayed in figure~\ref{doubleevent}. The 
level of gray represents the result of the cuts. The dark pixels are accepted NaF photons. 
The medium gray pixels are accepted AGL photons, while the light gray pixels have been 
rejected as background.
The events were then validated after the following condition on the consistency of the 
velocity measurements was fulfilled :

$$\frac{\beta_{NAF}-\beta_{AGL}}{\frac{1}{2}(\beta_{NAF}+\beta_{AGL})}\le 5\,
\sigma_{\beta_{NAF}}\approx 0.05$$
 
\par 
The distributions obtained from the analysis of this double $\theta_c$ 
measurement are displayed in figure \ref{doublethetac}. The dark hatched histogram located 
around $\theta_c=15^{\circ}$  is the AGL photon distribution, while NaF photon angles are 
standing as expected around $\theta_c=41^{\circ}$. The AGL $\theta_c$ distribution is distorted 
due to the small ring size (of the order of the pixel pitch) that generates a double peaking 
distribution, depending on the position of passage of the particle on the detector (see section 
\ref{DEADAREA}). This effect was also observed 
for beam tests with NaF for low $\beta$ ions (small ring size). The upper tail of the AGL 
distribution, also visible in figure~\ref{comparemethode}(a), is a consequence of Rayleigh 
scattering in this radiator.

Figure \ref{doublebeta} shows the result of the double $\beta$ measurement in the momentum 
range above the AGL threshold. It is seen that the velocity measurement on the bottom 
histogram with $\sigma_{\beta}\approx \sigma_{\beta\,NAF} \oplus \sigma_{\beta\, AGL}$ 
is slightly improved with respect to the individual measurements. The mean $\beta$ value 
slightly larger than one comes from the inaccuracy on the mechanical setting of the drift 
distance which ultimately translates into the observed overestimate.
 
\subsection{Albedo particle rejection}\label{ALBED}

\begin{figure}[htb]
 \begin{center}
 \includegraphics[scale=0.3,angle=0]{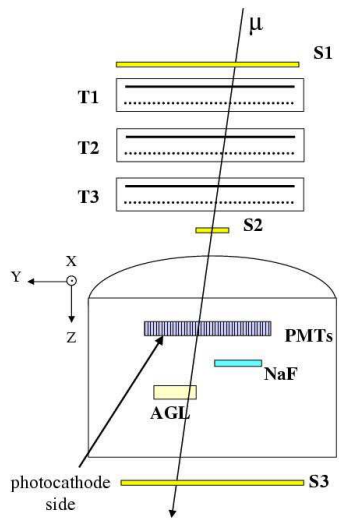}
 \end{center}
 \vspace{-1cm}
 \caption {\sl\small Schematic view of the experimental setup used for the Albedo tests.}
 \label{setup_albedo}
\end{figure}
%
\begin{figure}[htb]
 \begin{center}
 \includegraphics[scale=0.3,angle=0]{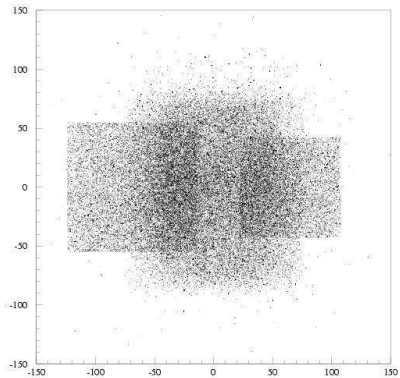}
 \end{center}
 \caption {\sl\small Distribution of reconsructed CR track intercepts on the radiator plane 
for particles producing one hit or more on the detector (see text). }
 \label{image_radiateur}
\end{figure}

Particles can enter the AMS spectrometer detector either from the top or from the bottom 
(Albedo particles). Since a particle entering from the bottom fakes an antiparticle entering 
from the top, it is clear that, disregarding the rejection power provided by the other AMS
detectors, the capability of the RICH to discriminate these two types of events, must be 
evaluated. Albedo particles are not expected to generate a response from the counter. 
However, some \cer light is produced, and then can be detected, and good events can be faked by 
unfortunate combinations of random hits. This issue has been adressed experimentally with 
the prototype. The results of the study have been reported in \cite{isola2000}. They are 
summarized here for convenience. 
\par
In the Albedo CR configuration, the prototype was set upside down, the photocathodes facing 
the ground. CRs entering from the top are thus equivalent to albedo particles for the RICH 
in orbit. Two radiators were tested in a single run: 
$10 \mbox{mm}$ NaF, d=70 mm, and  $3\times 11 \mbox{mm}$ AGL, d=157~mm. they were arranged 
as shown on  figure~\ref{setup_albedo}. 
\par 
Albedo data were taken with CRs for 14 days. The analysis showed that some \cer photons 
could reach the imager from both radiators. Figure \ref{image_radiateur} shows the 
distribution of the CR track intersections with the radiator planes, with the minimal 
condition of at least 1 hit on the detector required. The profiles of the NaF and AGL 
radiator tiles are clearly seen on the right and left of the figure, respectively. 
The rectangular shape visible at the center is an image of the global counter acceptance 
to CRs. It was generated by the fired PMT located on the track. 
This photon yield can generate events with multiplicity$>$1 that could fake \cer patterns. 
Among 22209 Albedo NaF events, 11 went through all cuts, whereas none of the 27104 AGL 
events could make it (see table \ref{tab_Albedo}). This difference is an acceptance effect : 
$\beta=1$ rings from AGL are smaller than those from NaF. Therefore fake rings have a larger 
probability to occur with NaF than with AGL.
\par 
The results could be well accounted for assuming that 
background hits are randomly distributed on the imager \cite{isola2000}. This successful 
interpretation enables us to estimate the rejection power for Albedo particles in the 
AMS RICH. For the NaF radiator, it is estimated to be $R\sim 10^{4}$ when requiring 3 hit 
minimum in the \cer cluster.  
Requiring a larger multiplicity improves the rejection efficiency, each extra unit enhancing 
the rejection power by a factor $\approx 50$. For the same requirements, the rejection power
obtained with the AGL 1.025 radiator is very good $R\approx 10^6$. However, Albedo particles 
have momenta typically $P \lesssim 5$~GeV/c per nucleon, and the threshold of this radiator 
$P\approx 4.1$ GeV/c/nuleon , is of marginal interest for the rejection purpose. \\
Therefore, the RICH could contribute efficiently to the Albedo particle rejection.

\begin{table} 
\begin{center} 
\begin{tabular}{l|cc}
\hline
\hline 
Radiator & NaF & AGL \\
Drift distance [mm] & 70.2 & 157.4 \\
$R_c$ [mm] & 130 & 51 \\
Statistics & 22209 & 27104\\
Through cuts & 11 & 0\\
Rejection power& 2019& $>27104$\\
Probability& $4.95\times10^{-4}$& $<3.68.10^{-5}$\\
\hline
\hline
\end{tabular}
\caption{\sl\small Experimental results for the Albedo particle detection for the two radiators.
        $R_c$ is the ring radius expected for $\beta=1$ particles. Statistics is the whole number of events with 
	a valid reconstructed trajectory falling in the acceptance of the detector array. Through cuts is 
	the number of events that faked a \cer ring (see text). The Rejection Power is 
	the ratio of the full counting statistics over the number of validated events passing
        the cuts. Probability is the inverse of the
	Rejection power that gives the probability to mistake an albedo particle as a 
normal particle in the RICH.}
\label{tab_Albedo}
\end{center}
\end{table}   

%
\section{Monte Carlo simulation}\label{SIMU}


\begin{figure}[hhh]
\begin{center}
\begin{tabular}{cc}
\includegraphics[scale=0.4,angle=0]{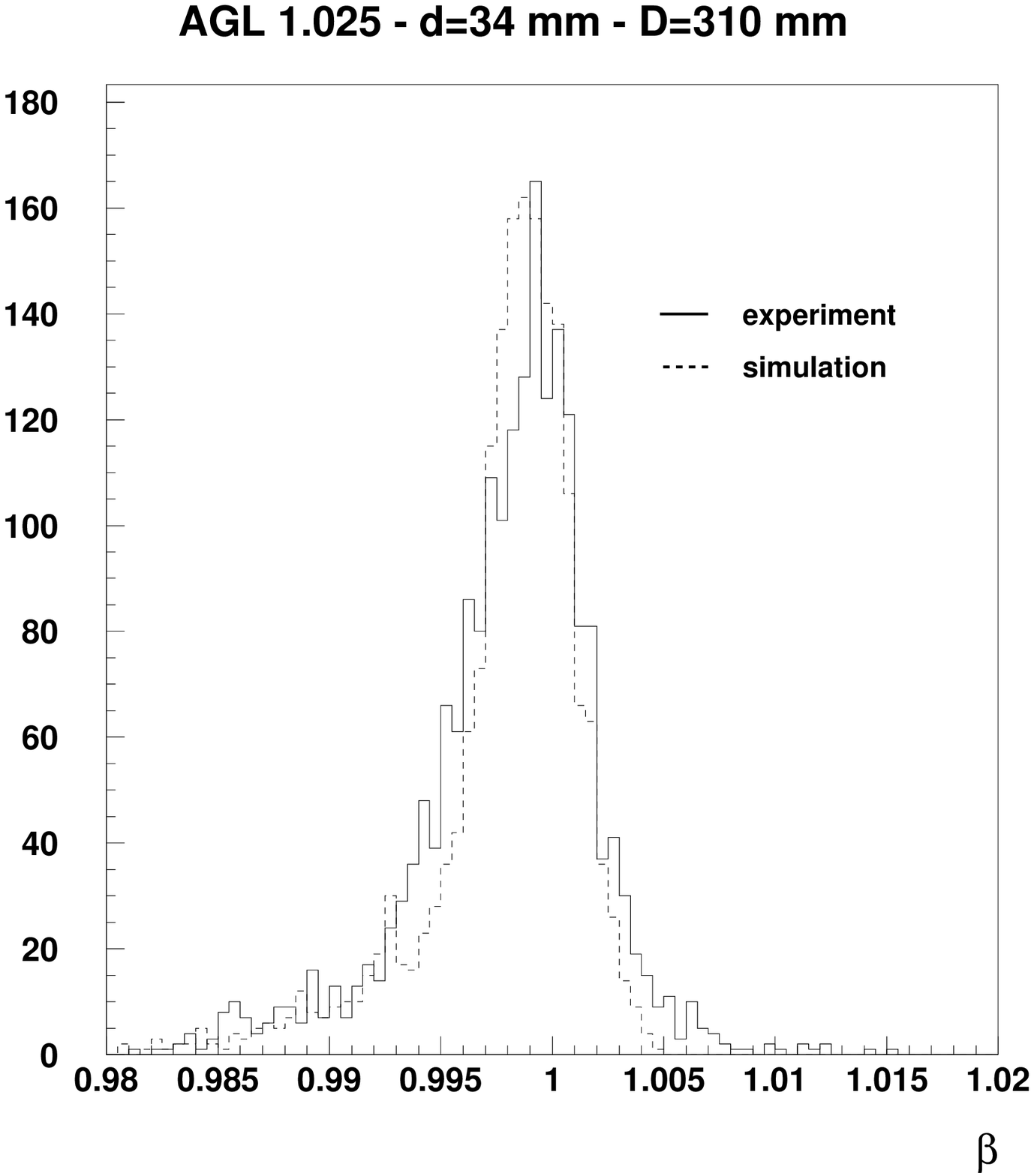} & 
\includegraphics[scale=0.4,angle=0]{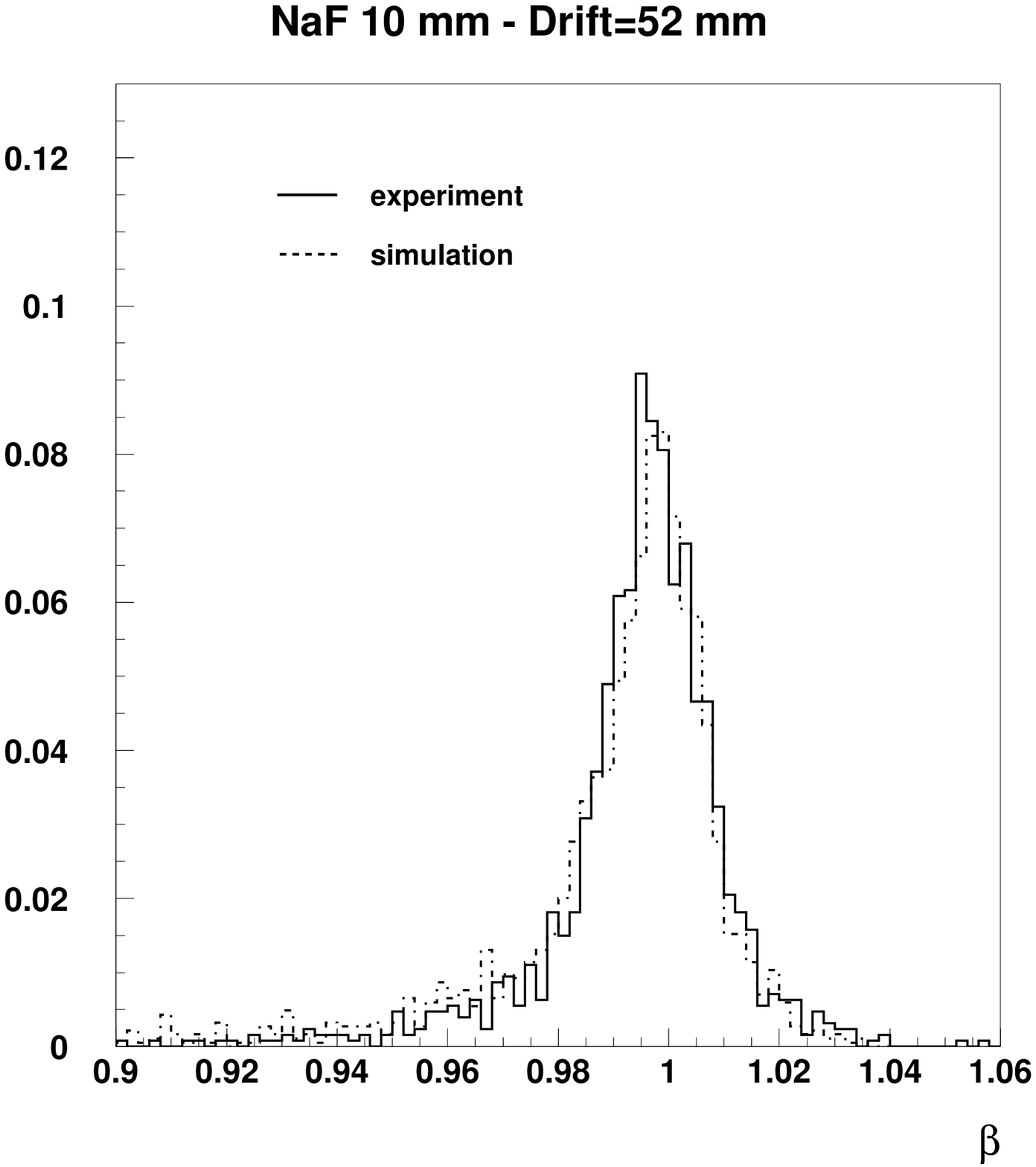} \\

    (a) & (b) \\  
  \end{tabular}
  \end{center}
  \caption{\em\small Comparison of the simulation results (dashed lines) with the 
experimental data for cosmic rays (muons, solid line) for AGL radiator $n=1.025$ (a) and  
NaF radiator (b). \label{comp_simu_exp}} 
  \end{figure}

The experimental CR data have been compared to simulation results performed with an adapted 
version of the code used in the first simulation study of the RICH \cite{SIMU}, using the 
same event reconstruction procedure. No background contribution was included in this 
simulation. 
\par 
The muon flux at ground level was described as in \cite{cosmicray}. 
The small proton flux ($\approx$10\% of muon flux) was included and the electron flux was 
neglected (\cite{BKLET}). The relevant PMT characteristics were taken into 
account : collection efficiency of first dynode assumed to be $0.9$, quantum efficiency of 
photocathode according to the technical datasheet of the manufacturer, SPE resolution 
taken the same for all PMTs as the mean value from calibration measurements 
(see section \ref{QGAIN}). The uncertainty on the reconstruced coordinate of the particle hit 
in the mid plane of the radiator was modelised by a gaussian of $2$~$\sigma$ in both X and Y 
direction, the value of $\sigma$ including both spatial resolution of the MWPCs and 
multiple scattering in the radiator.
\par 
The result for the NaF radiator is in good agreement with the experimental CR data 
(see table~\ref{tab:simu_vs_exp} and figure~\ref{comp_simu_exp}(b)). The velocity resolution 
is reproduced to within a few \%, as well as the hit pixel multiplicity. 
This point is important since it validates the predictions of the simulation program. 
The origin of this good agreement is mainly due to the fact that the optical properties of 
the radiator are very well known (chromatism, transparency...).  
\par 
The AGL radiator light yield were less straightforward to modelize because their optical 
properties were not as well known as for NaF, and because of the secondary effects in the 
light transmission (absorption and Rayleigh scattering). 
The AGL clarity coefficients $C\approx 10^{-2}$ $\mu$~m~$^{4}$~cm~$^{-1}$ \cite{MATSU} were 
available from the manufacturer for the n$<$1.05 samples. 
The AGL chromatic dispersion used was based on the scaling law $\frac{\delta (n-1)}{n-1}=cte$ 
using the measured values of silica (see appendix and \cite{SIMU}). This approximation has 
been recently shown to be in excellent agreement with the data \cite{VI01}. The results 
obtained however, are not good as seen in table~\ref{tab:simu_vs_exp} and figure~\ref{comp_simu_exp}(a), since the simulated 
resolution is significantly better than measured, although the simulated hit pixel 
multiplicity is in reasonable agreement with the data. No satisfactory explanation has been 
found for this discrepancy.
  
 \begin{table}
 \begin{center}
  \begin{tabular}{cccc|c|c|c|c} 
 \hline
 \hline
  radiator & $\langle n \rangle$ & $d$  & $D$  & \multicolumn{2}{c|}{ $\frac{\delta \beta}{\beta}\times 10^{3}$ } &\multicolumn{2}{c}{ $N_{pix}$}\\
\cline{5-8}
            &                          &      [mm] & [mm]    & sim.  & exp.  & sim.& exp.\\
 \hline
    NaF      & 1.332 & 10 & 52  & 8.9 & 8.8 & 7.5 &7.7 \\
   AGL      & 1.14  & 13   & 119 & 3.6 & 7.9  & 5.4 & 4.2 \\
   AGL      & 1.05  & 25   & 220 & 2.7 & 4.7 & 5.3 &  5.8  \\
   AGL      & 1.035 & 33   & 245 & 2.1 &3.5 &6.5&4.7  \\
   AGL      & 1.025 & 34.5 & 310.3 &1.9 &2.8 &4.4& 4.0 \\
   \hline
   \hline
   \end{tabular}
\caption{\em Comparison of the experimental CR results with simulation. $\langle n \rangle$ 
is the mean refraction index of the radiator, $d$ the radiator thickness, $D$ the drift 
distance, and $ {(\frac{\delta \beta}{\beta})}$ the velocity resolution from the simulation 
(sim) and measured at $\beta\sim 1$ (exp), $N_{pix}$ being the number of fired pixels after 
cuts.
 \label{tab:simu_vs_exp}}
   \end{center} 
  \end{table}
%
\section{Measurements with beam particles}\label{BEAMTST}

\begin{figure}[hhh]
 \begin{center}

\includegraphics[scale=0.3,angle=0]{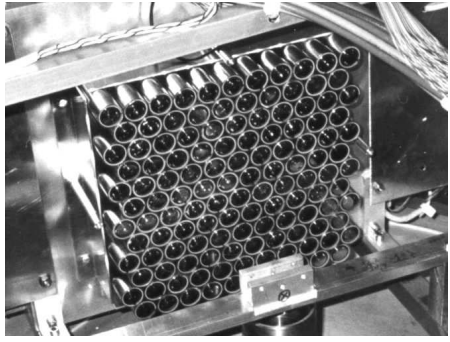}
 \end{center}
%
\begin{center}
\includegraphics[scale=0.5,angle=0]{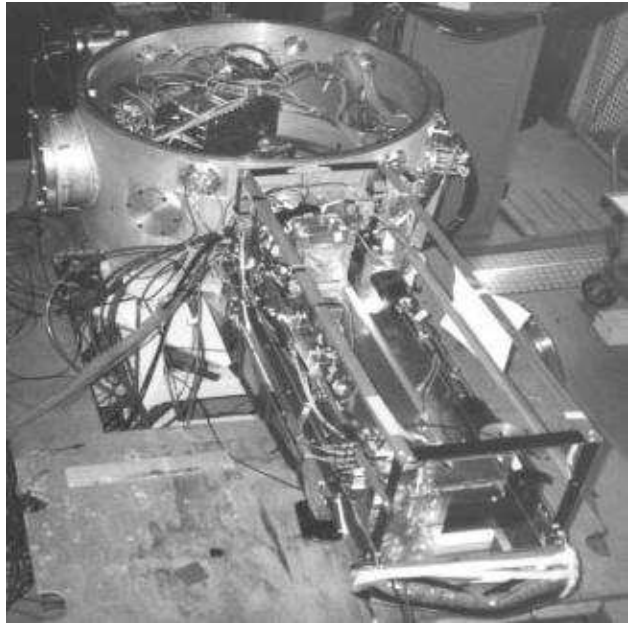}   
\caption{\em Closeup view of the PMT matrix (top) and general view of the setup used at 
GSI, showing the prototype in the vacuum chamber and the tracking and triggering detectors 
upstream.} 
\label{photo2}
  \end{center}
  \end{figure}

The prototype has been tested at the GSI/Darmstadt ion accelerator facility with $^{12}C$ 
beams with incident energies of 0.6, 0.8, 1, 1.2, and 1.4 Gev/nucleon. 
\par
In the test beam configuration, the detection plane was set vertical in the vacuum 
chamber, facing beam particles. The detector was placed on a movable arm around the fixed 
radiator holder. The radiator was placed at the center of the chamber and kept parallel to 
the detector surface, i.e., the two were rotating together, so that mesurements for 
different incident particle angles on the radiator could be achieved easily. The radiator 
could also be placed further upstream if necessary. The incident beam particle angles on 
the radiator could be varied between 0 and 45 deg. Two small MWPCs placed upstream of 
the chamber were used to define the incident trajectory, and a set of three small area
(typically $10\times10$ cm$^2$) scintillators framing the MWPCs along the beam line, 
provided the event trigger. The beam MWPCs were mainly used for providing the transverse 
coordinate of the particle hit on the detector matrix, for event reconstruction. 
Pictures of the matrix and setup on the beam line are shown on 
figure~\ref{photo2}. 

Only the NaF radiator could be tested in this experiment since the maximum beam velocity of 
the accelerator was below the \cer threshold for the aerogel materials considered 
(see table~\ref{DIMS}). The same NaF tiles were used as for the cosmic ray 
measurements. 
\par
Beam particles with different masses were obtained by placing a fragmentation target (a 
beam monitoring quartz was used to this purpose) upstream of the magnetic dipole analyzer. 
Fragments with atomic mass from 1 to 12 could then be obtained, with 
momenta defined by the field setting of the analyser. For a given beam energy, the various 
fragment mass and momenta could be obtain with a few bins of rigidity. With this method a 
set of data on nuclei over a range of mass and charge could be obtained to test the response 
of the prototype for each incident energy. Very low beam intensities were used, typically 
well below 10$^3$ particles.s$^{-1}$, with very low angular divergence $<\theta>\approx 
1$~mrd, and a size about 1x1~cm$^2$ at the target. Higher intensities could be used when 
fragments with A/Z different from the incident beam value were selected, the primary beam 
particles being off the detector area. 
The measurements were performed during a three days run on April 1998. Figure
\ref{ano_gsi} shows an example of Cherenkov ring for a helium particle.

\subsection{Effects of the photodetector dead area}\label{DEADAREA}

\begin{figure}[htbp]
 \begin{center}

 \includegraphics[scale=0.4,angle=0]{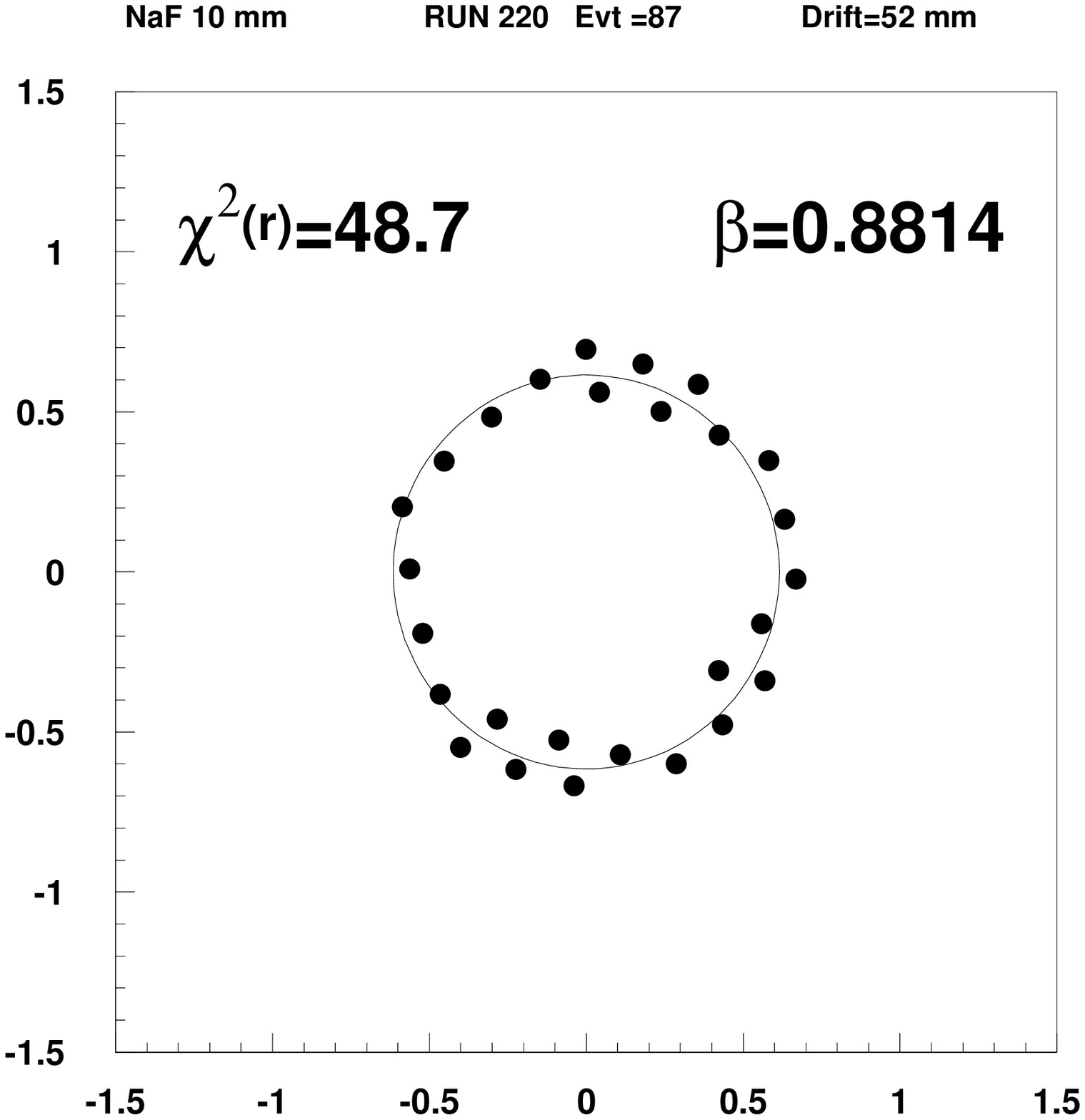}
 \end{center}
 \vspace{-1cm}
 \caption{\em Fit on the \cer ring produced by a Helium ion, reconstructed in the 
 reference frame of the trajectory.}
 \label{ano_gsi}
%
\begin{center}
\begin{tabular}{cc}
\includegraphics[scale=0.35,angle=0]{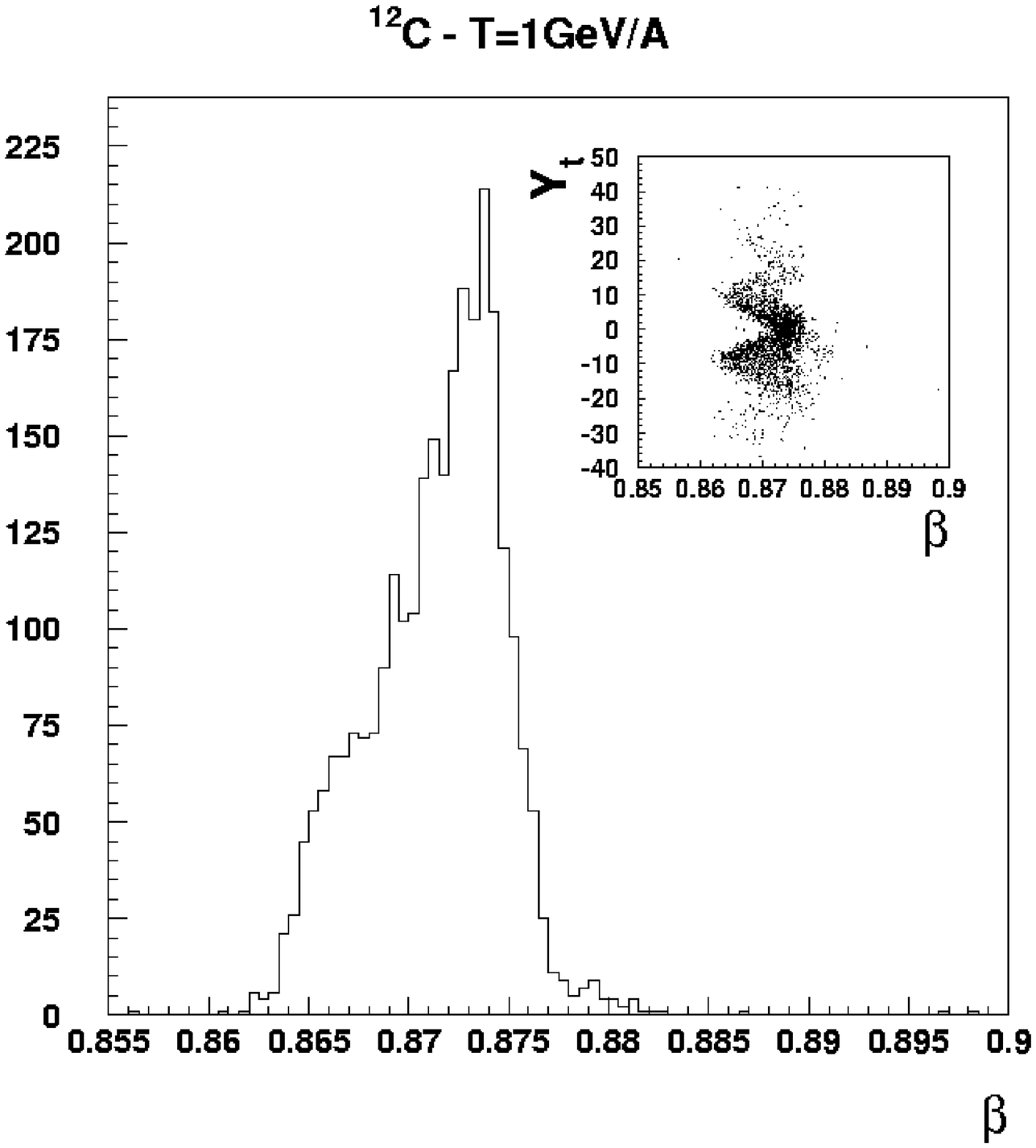} & 
\includegraphics[scale=0.35,angle=0]{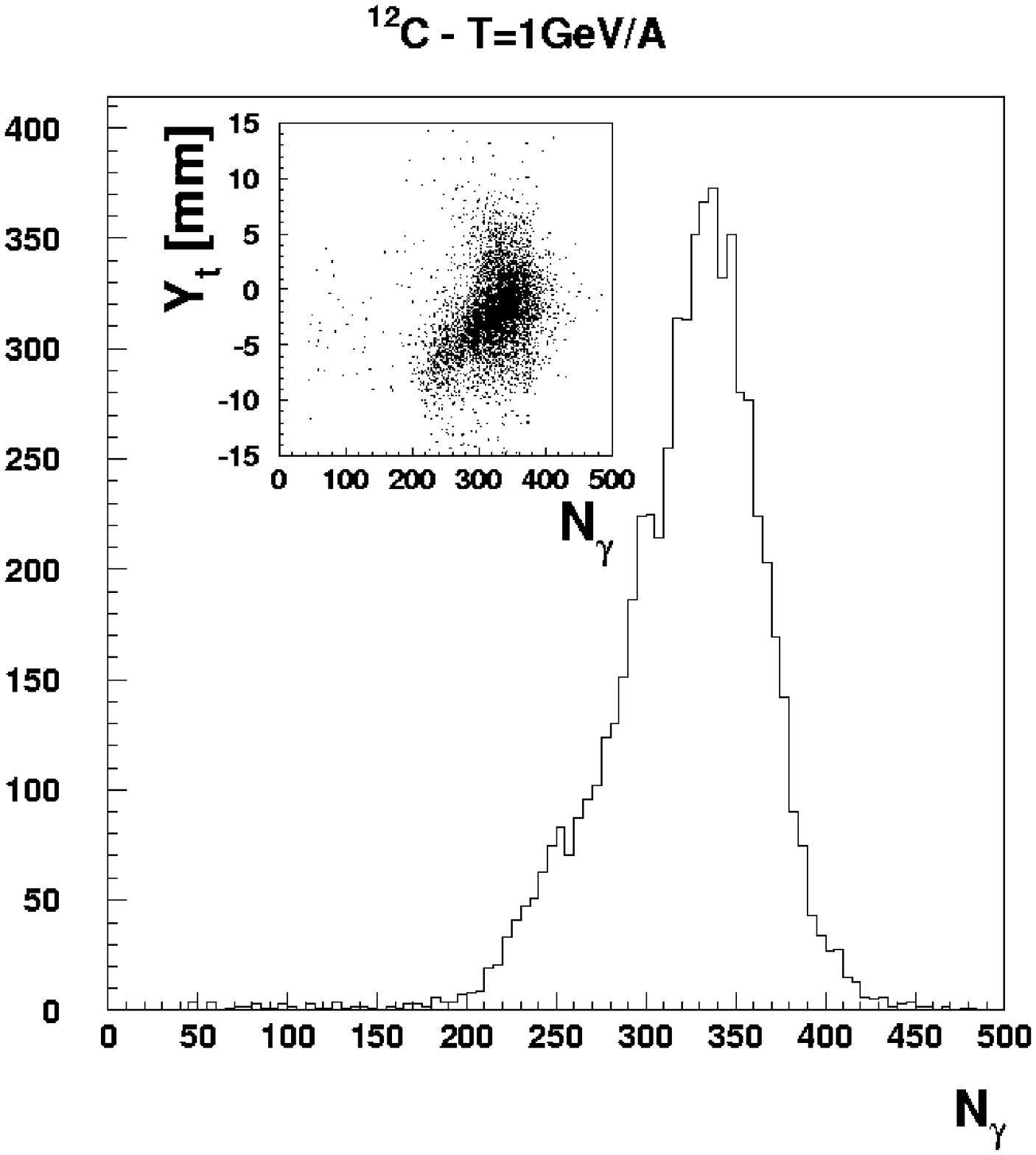}  \\
  \end{tabular}
  \end{center}
\vspace{-1cm}
  \caption{\em Distribution of reconstructed $\beta$ values (left) and number of 
  photoelectrons $N_{\gamma}$ (right) for $^{12}C$ beam particles at $T=1 $ GeV/A. 
  The inserts show the evolution of $\beta$ and $N_{\gamma}$  respectively, as a 
  fonction of the Y coordinate of the center of the reconstructed ring on the imager.
  \label{beta_gsi}} 
  \end{figure}

\begin{figure}[hbtp]
 \begin{center}
 \includegraphics[scale=0.4,angle=0]{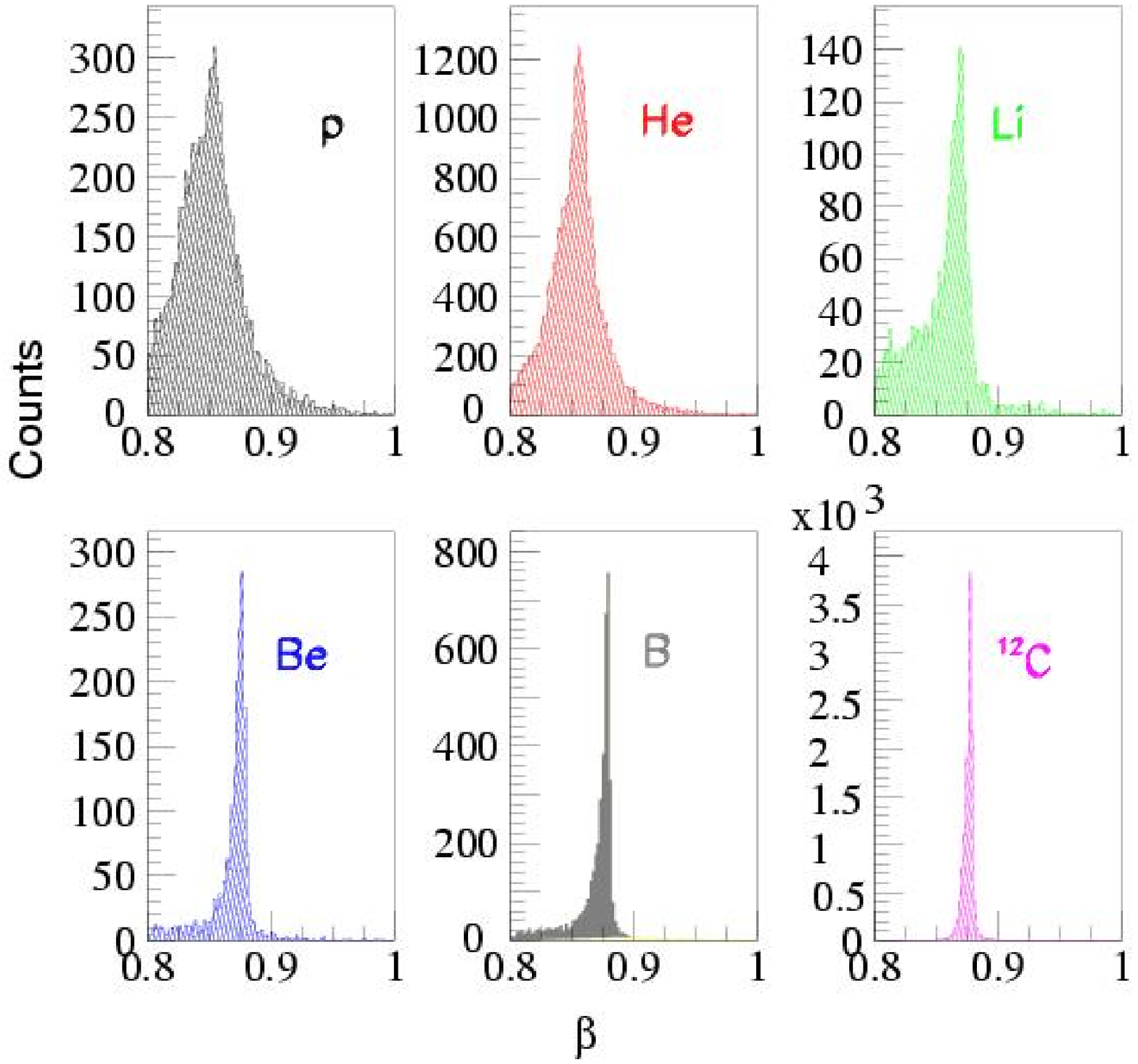}
 \end{center}
 \vspace{-1cm}
 \caption{\em Experimental velocity $\beta$ obtained with $T=1~GeV/A$ ${12}$C beam fragments 
 obtained as described in the text. The elements were identified separately by means of the 
 redundant of dE/dX measurement provided by the trigger scintillator counters \cite{THESETH}.
 The Z=6 incident beam particle spectrum was been taken from a separate run.}
 \label{betaions}
 \begin{center}
 \includegraphics[scale=0.4,angle=0]{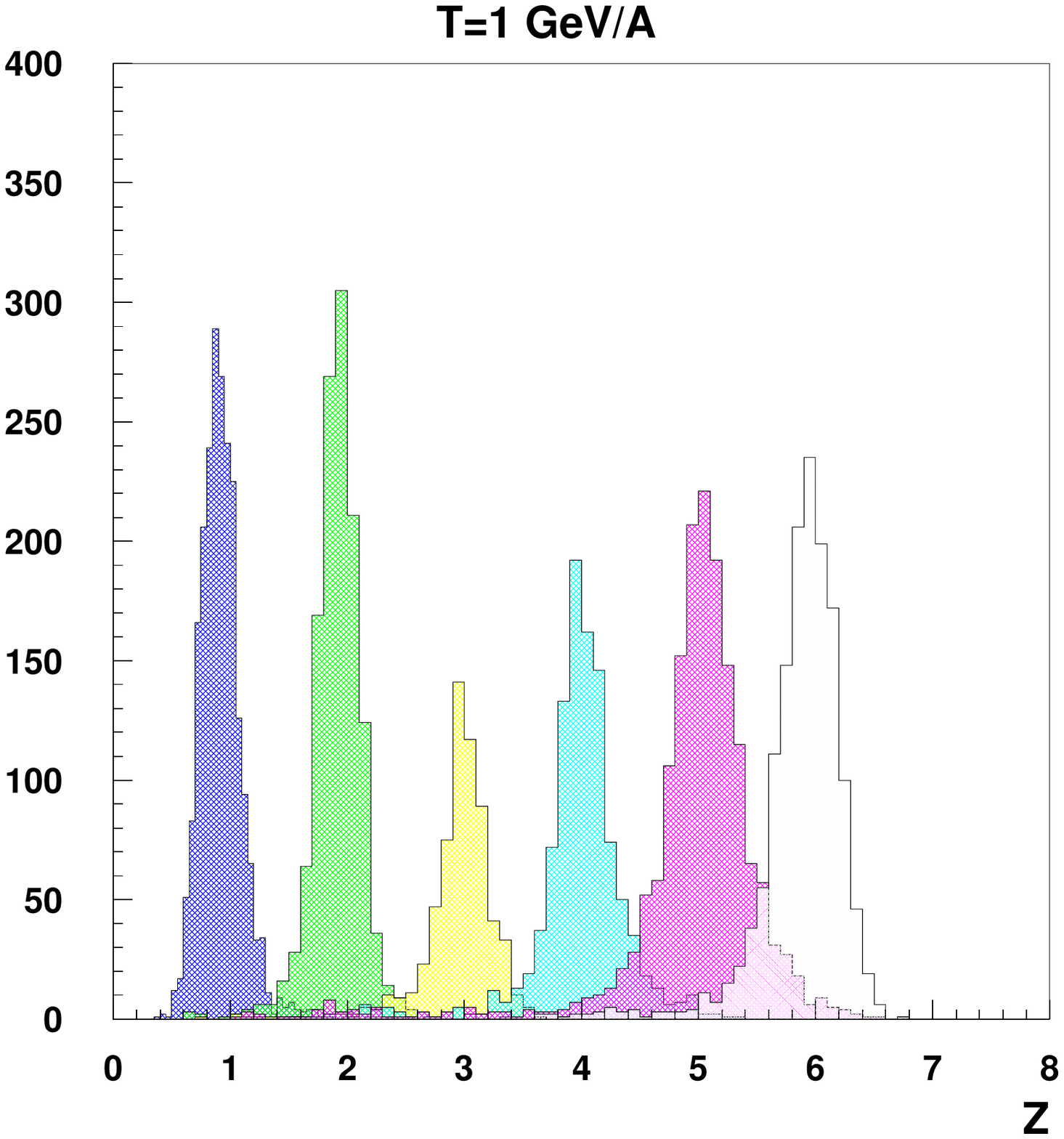}
 \end{center}
 \vspace{-1cm}
 \caption{\em Experimental Z reconstruction obtained with $T=1~GeV/A$ ${12}$C beam 
 fragments in the same conditions as in figure \ref{betaions}.}
 \label{z_gsi}
\end{figure}
%

The active area (packing fraction) of the photodetector was about 56\% of the total surface. 
This dead area had a very regular pattern which particular geometry generated some 
distortion effects in the event reconsruction. The main reason holds to the (radial) ring 
width and pattern cell dimension of the dead area being of the same order of magnitude. 
In this case, the overlap between the \cer ring and the active detector area depends on the 
ring position on the imager. This effect induces a systematic error which has been observed 
experimentally during beam tests. It generates a tail at low values in the distribution of 
the number of detected photoelectrons (figure~\ref{beta_gsi}(a)), and in the $\beta$ 
distribution (figure~\ref{beta_gsi}(b)).

This systematic error has been investigated and can be very well reproduced with a simple 
model \cite{THESETH}. It is also reproduced with the simulation program. It can be corrected
for by means of a simple algorithm which provides good results for the charge reconstruction. 
The correction on $\beta$ is more difficult to implement since the ring width depends 
on $\beta$ and is lower than the pixel size. Pixel size and packing fraction of the final 
imager however will largely exclude this effect.

\subsection{Particle velocity and charge reconstruction}\label{BEAMCH}

The data sample analyzed for $\beta$ and Z reconstruction included only normal incident 
particles, with \cer rings fully contained inside the detector area, then with no 
acceptance correction necessary. 

The final particle velocity reconstruction is illustrated on figure \ref{betaions}. The 
$\beta$ resolution was observed to improve with the increasing $Z$ of the particle, as 
expected from the \cer radiation law. The agreement with the expected $Z$ dependence 
according to the relation: 
$$[\frac{\delta\beta}{\beta}]_{Z}=\frac{1}{Z}[\frac{\delta \beta}{\beta}]_{Z=1}\;,$$ was fair 
to within the experimental uncertainties \cite{SIMU} as seen in table~\ref{ngamma_gsi}. 
Indeed, it can be seen that the quantity $Z [\frac{\delta\beta}{\beta}]_{_Z}$ is 
approximately constant, the larger excursion from the expected value for $Z$=1 and 2 being
mainly due to the distorsion effect quoted in the previous section. 
\par
The particle charge reconstruction was based on the $Z$ dependence of the \cer photon yield:

$$N_{\gamma}\propto \, Z^2\sin^2\theta_c $$

The Z separation depends practically (although it should not theoretically \cite{SIMU}) on 
the photon yield. It was typically 15 $PE$ for $\beta=0.85$ protons. The hit 
rejection cuts used for this reconstruction were softer than for the $\beta$ reconstruction, 
in order to avoid rejecting good hits. Only those hits located on the reconstructed 
trajectory were discarded. The effect on the $Z$ resolution would be significant with an AGL 
radiator because of Rayleigh scattering of \cer photons. It is small however for a NaF 
radiator.
Table~\ref{ngamma_gsi} summarizes the results and the Z distribution histogram obtained is 
shown on figure~\ref{z_gsi}. The light ions with $Z\le 3$ are well separated with 
$\frac{\Delta Z}{\sigma_{Z}}\approx 5$. For heavier ions however, the separation degrades 
down to $3-4$ for $Z=3,4,5$. This effect seems to be due to the onset of a saturation of 
both the PMTs above $\approx 20$ PE per pixel and (some of) the ADC electronics, in account 
of the high PMT gain used.
This effect should disappear with smaller pixel size PMTs, as it is foreseen for the AMS 
RICH. 

\begin{table}
\begin{center}
\begin{tabular}{ccccccccc}
\hline
\hline
element & $\beta$ & $\frac{\delta\beta}{\beta}\times 10^3$  
                            & $N_{pix.}$ & $N_{\gamma}$ & $\sigma_{\gamma}$& $Z$ 
                                            & $\sigma_{Z}$
                                                       & $\frac{\Delta Z}{\sigma_{Z}}$\\
\hline
   p  & 0.85  & 10.8& 7.75  & 14.8  & 7.5   & 0.9 & 0.15 &     \\
   He & 0.864 & 8.7 & 15.8  & 65.2  & 17.2  & 1.9 & 0.17 & 5.9 \\
   Li & .875  & 4.1 & 21.5  & 177.8 & 28.5  & 3.0 & 0.19 & 5.2 \\
   Be & .882  & 3.3 & 27.0  & 342.9 & 40.6  & 4.0 & 0.24 & 4.2 \\
   B  & .885  & 2.8 & 31.6  & 551.3 & 70.6  & 5.0 & 0.28 & 3.6 \\
   C  & .873  & 2.4 & 28    & 701   & 56.8  & 5.95& 0.24 & 4.2 \\
\hline
\hline
\end{tabular}
\caption{\em Reconstructed $\beta$ and $Z$ values for a $T=1$ GeV/A particle beam with 
normal incidence on the radiator. $N_{pix}$ is the mean hit pixel multiplicity, 
$N_\gamma$ the mean number of $PE$, $\sigma_{\gamma}$ the RMS of this latter distribution, 
$Z$ the reconstructed charge and $\sigma_{Z}$ the RMS of the Z distribution. 
$\frac{\Delta Z}{\sigma_{Z}}$ is the separation between two consecutive 
charges in RMS units (i.e., $\frac{1}{\sigma_{Z}}$).}
\label{ngamma_gsi}
   \end{center} 
  \end{table}

\section{Summary and conclusion}\label{DISC}
The study of a first generation prototype of proximity focussing RICH counter for the AMS 
experiment reported in this paper has allowed an end-to-end investigation of the technique:
Instrumental test of the detector components and electronics,  test of the reconstruction 
and background rejection algorithm, background measurement, and finally measurement of the 
counter resolution with different radiator samples using both incident cosmic rays
and beam ions with Z$<$6, casting the grounds of the future AMS RICH counter.
\par
The above work is being followed by a second generation prototype which incorporates 
the main features and elements of the final RICH design (flight model). It will be operated 
using the same instrumental peripheral environment as in the present work. This phase is 
being undertaken in collaboration between all the institutions involved in the effort on 
the RICH project \footnote{INFN Bologna, ISN Grenoble, LIP Lisbon, CIEMAT Madrid, U. 
Maryland, and IFUNAM Mexico}. \\
\par
\vspace{0.5cm}
{\bf\large Acknowledgements.} \\
The authors are very indebted to R~.Simon for his invaluable help during the data taking
at GSI. They are extremely grateful to M. Yokoyama (Matsushita), J. Favier (LAPP Annecy) and 
P. Fisher (MIT), for providing aerogel samples, and to B.~Ille (IPN Lyon) for making the 
set of MWPCs available to the authors. They are also indebted to R. Blanc, T.~Cabanel, 
G.~Gimon and M.~Marton for their contribution to the detector assembly, to A.~Garrigue, 
F.~Vezzu, and E.~Perbet, for their contribution to the mechanical study, to J.~Bouvier and 
O.~Rossetto for their help in the setting up of the electronics, and to Z. Ren for his help 
on the detector simulation. \\
One of the authors (A.M-R.) wishes to acknowledge the ISN hospitality and partial support of 
CONACYT and DGAPA-UNAM. \\
This work was made possible by a dedicated grant from the IN2P3/CNRS. 

\newpage
\begin{center}Appendix:
{\Large \bf  Refractive index and \cer radiator thickness}
\end{center}
This appendix briefly addresses the issue of the physical variables governing the refractive 
index and chromatic dispersion of materials. The implication on the thickness of the 
\cer radiators is discussed.
\par
The relationship between the refractive index  and the phsical of a medium is governed by 
the Lorentz-Lorenz ({\it L-L}) law, which can be expressed as 
\cite{LLL}:
\begin{equation}
\frac{n^2-1}{n^2+2}={\cal N}\alpha(\lambda)
\label{LoLoL}
\end{equation}
In this relation, $n$ is the refractive index of the material, {$\cal N$} the number density 
of particles in the medium, and $\alpha(\lambda)$ the dipole polarizability of the molecules
of the medium, i.e., their response function to electromagnetic driving forces. 
\par
For small values of $(n-1)$ it is straightforward to see that the above can be written as:
\begin{equation}
{n-1}\approx \frac{3}{2}{\cal N}\alpha(\lambda)
\label{LoLo2}
\end{equation}
Since ${\cal N}$ can be expressed in terms of the mass density $\rho$ of the medium 
($\rho={\cal\frac{N}{V} A})$, with ${\cal A}$ the molar mass of the material, 
and {$\cal V$} the Avogadro number), one has the relation of proportionality: \\
\begin{equation}
{n-1}\approx\rho\alpha(\lambda) 
\label{LoLo3}
\end{equation}
This simplified form of the {\it L-L} equation puts in evidence a few important properties 
of the refractive index of (transparent) materials: \\
1) - The quantity $(n-1)$ scales with the density $\rho$ of the material. Therefore, $(n-1)$
will change by approximately 3 orders of magnitude between the gas phase (under atmospheric
pressure) and the solid phase for a given element. \\
2) - The dependence of $(n-1)$ on the wave length $\lambda$ of the incident light is 
governed by the response function of the molecules of the medium to the corresponding 
electromagnetic perturbation.  
The relative variation of $(n-1)$ over a given range of $\lambda$ is thus given by the 
relative variation of the molecular response function $\alpha$: \\
\begin{equation}
\frac{\Delta(n-1)}{n-1}\approx\frac{\Delta\alpha}{\alpha}
\label{LoLo4}
\end{equation}
\par
Therefore, the scaling law $\frac{\Delta(n-1)}{n-1}\approx constant$ holds rather strictly 
to within the validity of the approximation for a given material. 
\par
The derivative of equation \ref{LoLoL} can be evaluated rigorously however, leading to: \\
\begin{equation}
\frac{2n}{(n^2+2)(n+1)}\frac{\Delta(n-1)}{n-1}=\frac{\Delta\alpha}{\alpha} 
\label{LoLo5}
\end{equation}
The evaluation of the term multiplying the quantity $\frac{\Delta(n-1)}{n-1}$ in this 
relation can be verified to be about constant, close to 0.3 for values of n between 1 
and 1.5. The approximation 
\begin{equation}
\frac{\Delta(n-1)}{n-1}\approx constant
\end{equation}
is then basically correct, although it is more accurate to use relation \ref{LoLo5}.
\par\noindent
3) - It is important to note that the chromatic dispersion of ($n-1$) also scales with the
matter density, i.e.:
\begin{equation}
\Delta(n-1)\approx\rho\Delta\alpha(\lambda)
\end{equation}
$\Delta\alpha(\lambda)$ being taken over some relevant range of $\lambda$. This explains 
in general why the chromatism of low density materials, like gas or aerogels, is much 
smaller than that of high density materials like crystals. This explains in particular 
why it is so for aerogels compared to quartz or fused silica, and it provides a way of 
estimating the chromatism of the former from the known dispersion law of the latter. \\
\par
{\bf Thickness of Radiator material} \\
The above discussion has straightforward implications for the thickness of the radiator 
material to be used for a RICH counter. This thickness can be expressed in terms of the 
\cer variables. 
The number of photons radiated is N$_{ph}$=N$_0$Lsin$^2\theta$, where N$_0$ is the quality
factor of the counter \cite{SIMU}, L the radiator thickness, and $\theta$ the \cer angle. 
One has therefore $L=\frac{N_{ph}}{N_0 sin^2\theta}$, or $L\approx\frac{N_{ph}}{2N_0<n-1>}$
for small values of $(n-1)$. Using relation \ref{LoLo3} above: 
$L\approx\frac{N_{ph}}{2N_0\rho<\alpha>}$, or
\begin{equation} 
\rho L=\frac{N_{ph}}{2N_0<\alpha>}
\label{RHOL}
\end{equation}
The quantity $\rho L$ is the thickness of the radiator in g/cm$^2$. It is seen that this 
quantity is constant for a given number of photons and for a given material. Although the 
quality factor can be somewhat different however for different values of $n$, this effect 
is small for refractive index not too much different, like between 1.02 and 1.1 in silica 
aerogels. 
With this restriction, relation \ref{RHOL} shows that the thickness of material to be used 
for a given number of photoelectrons does not depend on the mean refractive index of the 
material with the same molecular structure. For different materials the relation does not 
hold since the asymptotic value of $\alpha(\lambda)$ depends on the value of the first pole
of the dispersion law \cite{LLL}, which can differ by an order of magnitude from one material
to another.
%

\par


\begin{thebibliography}{99}

\bibitem{FRICH} J. Litt and R. Meunier, Ann. Rev. Nucl. Sci. 23(1973)1; 
\bibitem{YPS}   J. S\'eguinot and T. Ypsilantis, Nucl. Inst. and Meth. in Phys. A343(1994)30 
\bibitem{HYPO}  J. Ballon et al.,  Nucl. Inst. and Meth. in Phys. A338(1994)310
\bibitem{SIMU}  M. Bu\'enerd and Z. Ren, Nucl. Inst. and Meth. in Phys. A454(2000)476
\bibitem{AMS}   S. Ahlen et al., Nucl. Inst. and Meth. in Phys., A350(1994)351; 
                S.C.C Ting, Phys. Rep. 279(1997)203
\bibitem{BE10}  A. Bouchet et al., Nucl. Phys. A688(2000)417c; 
\bibitem{ANTEPROT} Z. Ren et al., Nucl. Instrum. Meth. in Phys. A433(1999)172; 
                   T. Thuillier et al.  Nucl. Instrum. Meth. in Phys. A 461(2001)278
\bibitem{THESETH} T. Thuillier, Thesis, Institut des Sciences Nucl\'eaires, 
                  Universit\'e J. Fourier, Grenoble, May 15, 2000, report ISN 00-47. 
\bibitem{BE94} E.H. Bellamy et al., Nucl. Inst. and Meth. in Phys., A339(1994)468
\bibitem{CA94} P. Carlson et al.,  Nucl. Inst. and Meth. in Phys., A349(1994)577
\bibitem{GA99}   L. Gallin-Martel, J. Pouxe, O. Rossetto, and P. Stassi, Nucl. Inst. and 
                 Meth. in Phys. A 433(1999)444; L. Gallin-Martel, J. Pouxe, and 
                 O. Rossetto, Proc of the IEEE conf, Toronto, November 1998, report 
                 ISN/97-26, Grenoble, April 1997.
\bibitem{NEWELEC} L. Gallin-Martel, J. Pouxe, and O. Rossetto, {\it A new front end 
                  electronics for the AMS RICH}, ISN Grenoble report ISN/99-30, April 1999, 
                  eprint physics/9812018
\bibitem{DAQ}    R. Duet, N. Borrome, H. Harrock, T. Tran-Kahn, P. Didelon and J. Navarre,
                 OASIS Data Acquisition system, Internal report, IPN Orsay, 1994; 
                 D.~Barancourt, G.~Barbier and B.~Meillon, ISN Grenoble, private 
                 communication.
\bibitem{isola2000} T. Thuillier et al., Nucl. Inst. and Meth. in Phys. A461(2000)278
\bibitem{CAPR}   P. Carlson et al.,Nucl. Inst. and Meth. in Phys. A349(1994)577
\bibitem{cosmicray} O.C. Allkofer, P.K.F. Grieder, {\em Cosmic Rays on Earth}, Physics Data, 
                    ISSN 0344-8401.
\bibitem{BKLET}  Review of Particle Physics, Eur. Phys. J., C15(2000)150  
\bibitem{FI94} D.E. Fields et al., Nucl. Inst. and Meth. in Phys., A349(1994)431
\bibitem{DE97} R. De Leo et al., Nucl. Inst. and Meth. in Phys., A401(1997)187
\bibitem{AS98} Y. Asaoka et al.,  Nucl. Inst. and Meth. in Phys., A416(1998)236
\bibitem{GO99} A. Gougas et al., Nucl. Inst. and Meth. in Phys., A421(1999)249
\bibitem{AS00} E. Aschenauer et al., Nucl. Inst. and Meth. in Phys., A440(2000)338
\bibitem{DE01} R. De Leo et al., Nucl. Inst. and Meth. in Phys., A457(2001)52
\bibitem{NA98}   E.~Nappi, Aerogel and its applications to RICH detectors, 
                 Conf. on advanced technology in particle physics, Como 96,
                 Nucl. Phys. B., 61B(1998)270
\bibitem{VI01}   M.F. Villoro et al., Nucl. Instrum. Meth. in Phys. A, in press
\bibitem{MATSU}  Matsushita Electric Works Ltd, Osaka, Japan 
\bibitem{ATC}    D. Barancourt et al., Nucl. Instrum. Meth. in Phys. A 465(2001)306 
\bibitem{AGING}  M. Bu\'enerd ans T. Thuillier, AMS Note 99-11-04 and ISN Grenoble report
                 99-122. See also, J. Favier, R. Kossakovski and J.P. Vialle, AMS note 
                 2001-03-05
\bibitem{LLL} M. Born and A. Wolf, Principle of Optics, Pergamon, 1975.

%
\end{thebibliography}
\end{document}